# Fractional Orbital Bombardment Systems

**David Wright**[1]
Laboratory for Nuclear Security and Policy, Massachusetts Institute of Technology
Cambridge, MA, USA
https://orcid.org/0009-0001-5958-0172

**Steve Fetter**
School of Public Policy, University of Maryland
College Park, MD, USA

We analyze Fractional Orbital Bombardment Systems (FOBS) as a means of delivering nuclear weapons by placing warheads into partial orbits around the Earth, and show that they do not represent a game-changing advance in technology or capability. We consider potential uses of such a weapon, including evading missile defenses and delivering attacks with short warning time. FOBS could underfly current exoatmospheric defenses, but the required payload reduction may not justify its use if other countermeasures are deemed effective. If a FOBS attack could be successfully disguised as a satellite launch, the warning time could be short enough to raise fears of a decapitating strike. The development of such a capability could lead countries to shorten decision-making timelines in a crisis, which would be destabilizing. The US can take steps to ensure that very-short-warning attacks would not succeed, thereby enhancing deterrence during a crisis.

Fractional Orbital Bombardment Systems (FOBS) are weapons that are launched into orbit around the Earth and then deorbited toward a ground target before the weapons have completed a full orbit. The concept dates to the early years of the Cold War. While no FOBS have been deployed since the 1980s, a Chinese test in 2021 of a system that has been described as a modern FOBS has again called attention to this issue.

Our analysis shows that FOBS do not represent a game-changing advance in technology or capability. We show that they could be used to evade US missile defenses by delivering warheads at very low altitudes, although doing so would require a significant reduction in the payload that missiles could deliver. In principle, disguised FOBS attacks could also be used to launch very-short-warning attacks on US nuclear command and control centers, but such attacks would entail very high risk and would be considered only under the most dire circumstances, if a crisis had escalated to the point that the adversary believed a US nuclear strike was likely. The development of such a capability could lead countries to shorten decision-making timelines in a crisis, which would be destabilizing. The United States can take steps to ensure that very-short-warning attacks would not succeed, thereby enhancing deterrence during a severe crisis.

## 1. History

The concept of FOBS originated in the early 1960s as part of an interest in basing nuclear weapons in orbit, which would give weapons unlimited range as well as short warning because the time for reentry after giving a command to deorbit might only be a few minutes. Because the reentry trajectory would be short compared to that of a long-range missile, this was also seen as a way of potentially achieving greater accuracy.[2]

The idea of orbital basing was soon abandoned in favor of nuclear-armed intercontinental and submarine-launched ballistic missiles, which afforded greater physical control and easier maintenance of nuclear weapons. Interest in FOBS then shifted to responding to the development of defenses against long-range ballistic missiles. At that time, flying on a partial orbit and attacking from an unexpected direction could avoid sensors the adversary had deployed to detect incoming warheads and guide defensive interceptors. The US had built missile defense radars scanning the northern horizon, because ballistic missiles launched from the Soviet Union would travel over the north pole to reach targets in the United States. By instead flying on a trajectory in the opposite direction and approaching the United States from the south, Soviet warheads could avoid these radars.

While ballistic missiles can launch warheads on a northern trajectory to reach targets at ranges of roughly 10,000 km, warheads on a southern trajectory to the same targets would need to fly about 30,000 km, which is too far to be reached by a ballistic missile. The idea behind FOBS was to instead launch warheads into orbit and equip them with a thruster that would kick them out of orbit to reenter over a target before they had completed a full orbit.

The 1967 Outer Space Treaty requires that "States Parties to the Treaty undertake not to place in orbit around the earth any objects carrying nuclear weapons or any other kinds of weapons of mass destruction."[3] FOBS weapons were seen as avoiding this restriction because they traveled less than a full orbit.

The US stopped its studies of orbital bombardment by 1965, but the Soviet Union continued developing three FOBS programs.[4] The Soviets selected what became the OGCh ("Orbital Payload") system (officially, an R-36-O missile with OGCh payload), which was based on the Soviet R-36 ICBM (called SS-9 Scarp by NATO) and carried a megaton-class nuclear warhead.

The Soviet Union conducted flight tests in 1966-71, and the system was declared operational in 1968. Some tests were conducted at altitudes less than 200 km.[5] A US Air Force study in 1976 estimated that a FOBS warhead's deorbit thruster would burn for about a minute to start deorbit, and that reentry would require an additional 1.5-2 minutes,[6] providing the country under attack with only about three minutes of warning from the start of deorbit. In response to the Soviet program and to close off the option of an undetected attack from the south, the US built a large south-facing radar at Eglin Air Force Base in Florida, which became operational in 1969 and is still operating.[7]

The Soviet Union began dismantling its FOBS in 1982. The 1972 Anti-Ballistic Missile (ABM) Treaty, which limited US and Soviet missile defenses, removed a key motivation for the system. Signature of the 1979 SALT II treaty, which prohibited developing, testing, or deploying orbiting weapons containing weapons of mass destruction, "including fractional orbital missiles," was also a factor, even though the treaty was never ratified by either country.

The FOBS issue resurfaced in a 2018 speech by Russian President Vladimir Putin, which described several new systems Russia was developing to counter missile defenses the US began building after it withdrew from the ABM Treaty in 2002. One such system was Sarmat, a large ICBM that Putin described as having the ability to "attack targets both via the North and South poles."[8] Putin showed an animation in which Sarmat was launched north towards United States on a normal ballistic trajectory, and south on what appeared to be a fractional orbit.

Concern about FOBS intensified in 2021 with China's launch of an object into orbit that reportedly circled the Earth and released a hypersonic glide vehicle that reentered the atmosphere and dove toward a target on the ground.[9] This led to speculation about what kind of weapon China was developing, whether it was intended to evade US missile defenses, and how large a threat it posed.

China's capability to carry out such a test was not a surprise. China had been developing a "space plane"—apparently similar to the X-37B space plane the US was developing a decade earlier—that could be launched into orbit on a rocket and then deorbited to return to Earth and land like an airplane. China has carried out four successful tests of that vehicle, with the first of these tests in September 2020.[10] These tests showed that, like the US, China has developed technology that could be used either for a space plane or for an orbital ground-attack weapon like FOBS.

This renewed discussion of FOBS raises several questions:

- What are the potential uses of FOBS-like weapons today, and what advantages would they provide over other weapons?
- What capabilities would those uses require, and what does that imply about the required technology?
- How much of a threat might the use of such weapons pose?

Below we use the following terminology:

- FOBS: the launch vehicle and payload (FOBS weapon) that is placed in orbit
- FOBS weapon: the reentry vehicles(s), deorbit boosters(s), and guidance system
- Reentry vehicle (RV): the nuclear warhead, arming and fuzing system, and protective aeroshell.

## 2. Potential uses of FOBS

An official 1996 history of the Soviet missile program listed what were seen as potential advantages of FOBS, the primary one of which was evading a US missile defense system.[11] Evading missile defenses is still cited today as the main purpose of FOBS. As we discuss below, there are costs to using it as a countermeasure to defenses, and whether it is desirable depends on other types of countermeasures that are available and how effective the defense is expected to be.

A second potential use of a FOBS-like system is to deliver a small number of warheads with very short warning time, which might be the case for a "decapitation" attack that attempts to destroy key command centers and leaders early in a crisis. We discuss below how feasible such attacks may be and what uncertainties surround them.

### 2.1 Evading defenses

Developing effective ways to evade missile defenses was a greater concern when the US and Soviet Union were building nuclear-tipped interceptors able to destroy warheads at distances of several kilometers. In contrast, current exo-atmospheric interceptors must collide with a warhead to destroy it, and these "hit-to-kill" defenses are vulnerable to a range of countermeasures that can be developed by any country capable of producing a long-range missile and nuclear warhead.[12] Even so, a country might want to demonstrate other ways of evading missile defenses. In addition to alluding to a FOBS capability for Sarmat in his 2018 speech, Putin announced the development of the Avangard hypersonic glide vehicle and the intercontinental-range nuclear-powered Poseidon torpedo and Burevestnik cruise missile, all of which were portrayed as counters to US missile defense.[13]

As noted above, FOBS are frequently described as avoiding missile defenses by flying on trajectories passing over the south pole. Such warheads would avoid US early warning radars that track missiles coming from the north, and the warheads would not come within range of the Ground-Based Interceptors currently deployed in Alaska. But the warheads would be seen by the large south-facing radar located in Florida, and ship-based missile defense systems, such as the Aegis SM-3, could be positioned to engage warheads approaching from the south.

FOBS could, however, evade exo-atmospheric defenses even on northern trajectories by flying on low-altitude orbits at about 120 km. The infrared sensors on current US exo-atmospheric interceptors that guide the kill vehicle to its target cannot operate at these low altitudes because the high speed of the interceptors results in heating strong enough to blind these sensors.[14] The drag on a warhead at this altitude would not be problematic during a partial orbit.[15] A FOBS weapon on such a trajectory could therefore fly under the current US strategic missile defense system, the Ground-based Midcourse Defense (GMD).

Flying a warhead at 120-km altitude would also reduce the detection range by ground-based early warning radars from around 4,000 to 5,000 km for a long-range missile on a typical trajectory (MET) to perhaps 1,000 km. This would delay radar confirmation of the attack warning from early warning satellites, and would reduce the time the defense could track the threat, reducing the accuracy of its location predictions.

For this reason, if a country wanted to demonstrate additional ways to defeat exo-atmospheric missile defenses, FOBS might be an option. Whether it would provide significant advantages over other means of defeating missile defenses is unclear. FOBS imposes a substantial payload penalty both due to the extra velocity needed to insert a payload into orbit, and because the payload includes the booster needed to deorbit the warhead. For a range of 10,000 km, the same missile could deliver only a third to a half the mass of RVs by FOBS that it could on a standard ballistic trajectory.[16]

Whether FOBS makes sense as a missile defense countermeasure depends on estimates of the effectiveness of the defense—and confidence in that estimate. For example, if FOBS reduces the number of deliverable warheads by a factor of three but increases the probability of penetration to 100 percent, then FOBS should be favored when the expected ability for warheads to penetrate a defense is less than one-third.

To the extent that FOBS can penetrate national missile defenses more effectively and reliably than ballistic warheads and their associated countermeasures, they should strengthen deterrence and improve stability by providing additional confidence in second-strike retaliatory capabilities. This should, in turn, reduce the potential for arms racing and build-ups of offensive forces to offset missile defenses. On the other hand, if FOBS allow very-short-warning attacks, they could be destabilizing, as discussed next.

### 2.2 Short-warning-time attacks

Effective nuclear deterrence depends not only on nuclear forces that can survive an attack and penetrate missile and air defenses, but also on survivable command and control systems and procedures for detecting and assessing an attack and ordering a retaliatory strike. It is extremely difficult to imagine an attack that would destroy a large fraction of US nuclear forces, which are deployed in hundreds of hardened missile silos, on bombers that can be dispersed in a crisis, and on ballistic missile submarines patrolling undetected in the vast ocean. By contrast, the number of key command and control targets is small. Of particular concern is the possibility of a decapitation attack that would kill the president and other top civilian and military leadership that would formulate and order a retaliatory strike. The United States guards against this possibility in two main ways: with aircraft that allow the president and senior leadership to relocate and continue operations; and with procedures that allow the president to order a nuclear strike quickly, before incoming warheads detonate. With sufficient warning of an attack, officials also could relocate to hardened underground bunkers that could survive a nuclear explosion.[17]

As shown in Table 1, relatively little time is available to assess and respond to a nuclear attack. The total flight time of an ICBM on a traditional ballistic trajectory of 9,000 to 10,000 km is only 30 to 35 minutes. As little as 20 minutes would be available between alerting the president that an attack is underway and detonation of the incoming warheads. During this short period, the president and other senior leaders must be relocated and/or decide on a retaliatory response and transmit orders to deployed nuclear forces.

**Table 1**: US nuclear launch decision timeline for an ICBM or FOBS attack.[18]

| Event | ICBM | FOBS |
|---|---|---|
| Launch | 0 | |
| U.S. early-warning satellites detect hot plume from the booster; detection data transmitted to early warning teams at NORAD, STRATCOM, NMCC, and other facilities | 1 min | |
| Early-warning satellite data is analyzed to determine launch location, booster type, ascent profile | 1 to 4 min | |
| NORAD and STRATCOM early-warning teams must report initial confidence (no, medium, high) of nuclear missile threat to North America; NORAD and STRATCOM Commanders are notified and briefed on the apparent threat | 4 min | |
| If confidence is medium or high, POTUS is notified and reachable advisors are dialed into teleconference or video conference | 4 to 9 min | |
| STRATCOM Commander briefs POTUS on the threat and response options and makes recommendation | 9 to 10 min | |
| Early-warning radars detect ICBM warheads or FOBS; early warning teams confirm attack; ICBM target prediction updated | 10 min | 15 min |
| POTUS is updated, deliberates with advisors on response options | 10 to 17 min | |
| Early-warning satellites, radars detect FOBS deorbit booster burn | | 17 min |
| FOBS deorbit burn ends; FOBS target prediction updated | | 18 min |
| Warheads detonate on targets | 30 min | 20 min |

Launching warheads on low-altitude trajectories would give a short flight path compared to a typical ballistic trajectory over the same range, reducing the flight time from Russia to targets in the US to under 20 minutes.[19] As an example, consider an attack on Washington, DC, from the closest Russian ICBMs, which are deployed at the Vypolzovo missile field. The flight time to Washington on a 7,500 km minimum-energy trajectory (MET) from Vypolzovo would be about 30 minutes, and it would reach a maximum altitude of about 1,370 km. That time could be shortened to about 21 minutes by flying it on a less energy-efficient depressed trajectory (DT) with an apogee of about 300 km.

That time could be further shortened to about 18.5 minutes over this range by launching a FOBS into orbit at 120 km and then deorbiting over the target. In this case, the launcher would have to accelerate the warhead to the orbital speed of 7.8 km/s, compared to a burnout speed of 6.3 km/s for the MET and 7.2 km/s for the DT. As a result, the DT missile could carry only 80 percent of the payload of the original MET, and the FOBS could carry 45 percent of the MET payload, part of which would be taken up by the mass of the FOBS' deorbit boosters.[20] As noted above, however, a FOBS warhead orbiting at this altitude would have the advantage that it would be too low for current exo-atmospheric defenses to engage.

As with traditional ICBMs, the bright infrared signal from the FOBS missile booster would be detected and tracked by early warning satellites. If the booster was known to be associated with FOBS because of prior testing or other intelligence, the FOBS attack could be confirmed and the president notified as quickly as for a traditional ICBM launch. But the time available for relocation of senior leaders and/or deliberation and response would be cut in half, from roughly 20 to 10 minutes. This could affect the ability to relocate the president and senior leaders, and it would significantly decrease the time available to assess the attack and formulate an appropriate response.

Radar confirmation and target prediction would also come later for a FOBS attack than an attack on an MET, reducing confidence in the nature of the attack. An ICBM releases warheads on predictable ballistic trajectories shortly after the boost phase and the warheads rise to high altitudes before falling back to Earth, allowing accurate target prediction and confirmation of the attack by ground-based radars about 20 minutes before the warhead reaches the target. By contrast, FOBS weapons could travel at low altitude after the boost phase, reducing radar detection to less than 10 minutes, and targets could not be accurately predicted until the deorbit booster completed its burn, which could be as little as two minutes before the warhead reaches the target.

The United States has long required that the warning provided by satellites be confirmed by ground-based radars in order to have high confidence of an attack. The requirement that an attack be confirmed by independent systems using different physical principles is known as "dual phenomenology." The United States maintains the option of launching its silo-based ICBMs on warning of an attack, but under current doctrine that option can be exercised only if the attack is detected with both early-warning satellites and early-warning radars. The late radar detection of a FOBS attack would therefore limit how early the US could order a launch.

The importance of decreased warning time and delayed radar confirmation and target prediction, and the degree to which these factors might weaken deterrence and undermine strategic stability, is unclear. For the cases discussed above, enough time would be available for a US to launch a response, so adversaries could not be confident that a FOBS attack would prevent a retaliatory strike. In a crisis, response options might be pre-planned or the president or successors might be relocated to aircraft or facilities that could survive an attack, increasing the likelihood of retaliation.

As a result, these shortened delivery times could complicate retaliation, but are not short enough to give an attacker confidence that they could significantly reduce the possibility or scale of a retaliatory attack.

### 2.3 Very-short-warning-time attacks: Decapitation

If, however, warning time could be reduced under 10 minutes, such an attack could prevent the United States from being able to relocate the president or to formulate and execute a nuclear retaliatory response. A decapitation attack could be relatively small, targeting the senior officials involved in ordering a retaliatory strike, the aircraft that would be used to relocate these officials and serve as airborne command posts, and key command bunkers. The objective would be to:

- attack Washington, DC, before high-level officials, including the president, vice-president, secretary of defense, chairman of the Joint Chiefs of Staff, can relocate and/or transmit an attack order;
- attack US Strategic Command (STRATCOM) at Offutt Air Force Base near Omaha, NE, before the STRATCOM commander can depart on the Airborne National Command Post (Looking Glass); and
- destroy the other bases/aircraft that would allow officials to participate in an emergency decision conference or to formulate and transmit emergency action messages.

The highest priority targets might therefore include:

- White House (president and vice president)
- Pentagon (National Military Command Center)
- Joint Base Andrews (National Airborne Operations Center and Air Force One), 10 km southeast of Washington, DC
- Offutt Air Force Base (Strategic Command, Airborne National Command Post), Omaha, NE
- Raven Rock (alternate National Military Command Center), 100 km north northwest of Washington, DC

As few as five warheads (ten to allow for failures) would be required. If the attack could target two locations, they might be used to attack targets near Washington, DC, and Offutt Air Force Base. If the FOBS can divert additional warheads to widely dispersed targets, the attack might be expanded to include Tinker Air Force Base in Oklahoma (the base for the Navy's airborne command post) and Peterson Air Force Base and Cheyenne Mountain in Colorado (NORAD primary and alternate headquarters), although attacking these locations would likely require separate launchers.

Very short warning time might be possible if FOBS weapons could be disguised as satellites until they started to deorbit. This would give only a few minutes warning—too little to relocate senior officials or order a retaliatory strike. Such a possibility might tempt an adversary to risk a decapitating strike, particularly if it believed nuclear war was inevitable and that a decapitating strike was its best hope of survival.

One way to do this would be to station nuclear weapons in low-altitude orbits. These could be launched over time and disguised as satellites. Stationing nuclear weapons in space would be a violation of the Outer Space Treaty, but violations might not be detected. No technical means have been developed to determine whether a satellite is carrying a nuclear warhead, but there are proposals for such verification that could be implemented in the future to interrogate suspicious satellites.[21]

Human, signals, or other intelligence collection might reveal the existence of the on-orbit warheads, which would reduce their ability to execute a surprise attack and could lead to an attempt to shoot them down. There also is the difficulty of maintaining nuclear weapons in space for extended periods (especially replacing tritium, which decays with a half-life of 12 years) and recovering or disposing of the warheads at end of life. And even if a country was willing to put warheads in space, orbital motion would ensure that only a fraction would be in position for an attack at a given time, and the timing of the attack may have to be adjusted to ensure coverage of the targets.

A potentially more effective way to attack a small number of targets with short warning would be to launch a FOBS disguised as a satellite shortly before the intended attack. Satellite launches are a frequent and increasing occurrence, and are likely to continue during a crisis, both for commercial operations and to replace or augment military surveillance and reconnaissance capabilities. In 2025, the US, China, and Russia launched 193, 93, and 17 space launch vehicles, respectively—more than double the combined number in 2020.[22] These launches are detected by early warning satellites and quickly classified as satellite launches based on launch location (known space launch center), booster type (known space launch vehicle), and ascent profile, together with a corresponding NOTAM (notice to airmen) that warns air traffic of planned launches, published two or more days in advance of the launch.

To disguise a FOBS attack as a satellite launch, an adversary could mimic prior satellite launches, using the same space launch center and space launch vehicle and inserting the FOBS weapon into an orbit with inclination and altitude similar to a prior satellite orbit. Without prior human or signals intelligence about disguised FOBS operations, there is a good chance the weapon would be classified as a satellite and ignored as a potential threat. As we discuss below, however, the current practice of announcing satellite launches in advance and publishing NOTAMs would limit the attacker's ability to plan and launch such an attack on a short timescale while keeping it disguised as a civilian launch.

The satellite launcher could carry a payload that included several warheads. A deorbit booster could dispense the warheads toward various targets, or each warhead could be equipped with its own deorbit booster to direct it towards its target. If the United States believed the FOBS weapon was a satellite, the only warning of attack would come with the burn of the booster to start the deorbit, and it would take additional time for the United States to determine that an attack was underway. The deorbit time could then be only a few minutes, depending on the details of the orbit, target location, and thrust and burn time of the deorbit booster.

The targets of a disguised FOBS attack would be limited by the number of warheads that could be placed into orbit by one SLV, because launching multiple SLVs in a very short period of time would raise suspicions. This would limit the number and physical separation of targets that could be attacked in this way, as shown below.

A variation on this attack would be to launch the warheads on SLVs but allow the FOBS weapons to remain in orbit for several hours. This might allow an attacker to conduct two launches several hours apart, perhaps from different launch centers, with each orbit chosen to pass over its target after a given number of orbits, with the launches timed so that the weapons pass over their targets at the same time. We also consider this case below.

## 3. Details of FOBS Launch and Deorbit

To illustrate the issues related to FOBS, we consider hypothetical launches from Russia and China against targets near Washington, DC, and Omaha, NE. We assume the FOBS weapon is placed into a low-altitude, nearly circular orbit by either a space launch vehicle (SLV) or ballistic missile. The FOBS weapon consists of a nuclear warhead in a reentry vehicle (RV), attached to a deorbit booster that can provide the velocity change ($\Delta V$) to accelerate the weapon out of orbit toward a ground target.

The $\Delta V$ of the deorbit booster can be divided between increasing the vertical velocity, which will reduce the time to target, and producing a lateral velocity that will allow the weapon to travel to either side of the ground track of the original orbit. The lateral distance will be determined by the lateral velocity given to the RV and the deorbit time.

The deorbit booster's $\Delta V$ will also determine the amount of propellant required by the booster, and therefore the mass of the weapon. The mass increases exponentially with the $\Delta V$ of the booster, so achieving a large $\Delta V$ can lead to a very massive FOBS weapon. Our model of the FOBS booster, the equations used for calculating the booster mass, and the equations of motion used for calculating the trajectories, are given in the Appendix.

### 3.1 Launching the FOBS into orbit

We look at FOBS launches by both SLVs and ballistic missiles. For SLV launches, we consider launch sites in Russia and China that are used to launch satellites into roughly polar orbits. We examined the recent launch history at these sites to see what types of satellites have been launched, the altitude and inclination of their orbits, and what launch vehicles were used.

Because both countries have large numbers of deployed missiles, launching FOBS weapons on ballistic missiles could in principle allow many launches in a relatively short time. These launches could occur at known deployment sites, or at unknown locations from mobile launchers or submarines. Attacks disguised as satellite launches would be much more limited because only one SLV could be launched in a short period of time from a space-launch center without raising suspicions.

Two factors limit the capability of a missile or SLV to launch FOBS weapons. First, the total mass of the weapon will depend on the mass of the RV and the mass of propellant for deorbiting. Designing a weapon with a short reentry time and/or large lateral reach could require a large $\Delta V$, resulting in a large propellant mass. Second, FOBS weapons must be accelerated to a higher speed to place them in orbit than to fly on a long-range ballistic trajectory, which will reduce the payload that a missile booster can launch as a FOBS compared to on a ballistic trajectory. This is discussed in more detail below.

### 3.2 FOBS vehicle mass, deorbit time, and footprint

Figures 1 to 4 show the results of calculations for the mass of the FOBS weapon, the time of deorbit, and the deorbit footprint for two cases that we discuss in more detail later. These calculations assume no lift forces on the RV during reentry, but we discuss adding lift below. Here we assume the RV mass, $M_{RV}$, is 200 kg, but as shown in the Appendix the total FOBS weapon mass scales with the RV mass, so these results can be scaled for other values of $M_{RV}$.

Figures 1 and 3 show that the FOBS weapon mass increases rapidly as $\Delta V$ increases above about 2 km/s. The mass is important because it determines the launch capacity required to put the weapon in orbit.

Figures 2 and 4 show curves of possible impact points on the ground, which we call footprints, for a FOBS weapon that fires its deorbit booster at the point (0,0) and time $t = 0$. (As a shorthand throughout the paper, we refer to the endpoint of the trajectories as "impact" points, although the warheads would likely be detonated some distance above the ground.) These plots assume a given total value of $\Delta V$, which can be divided between creating downward and lateral velocities, and show the tradeoff between deorbit time and crossrange distance for the given orbit altitude and value of $\Delta V$. Impacting at points in the interior of the curves can be accomplished by delaying the ignition of the booster so that deorbit begins further along the orbit than in these figures. Below we show footprints superimposed on maps of potential target areas.

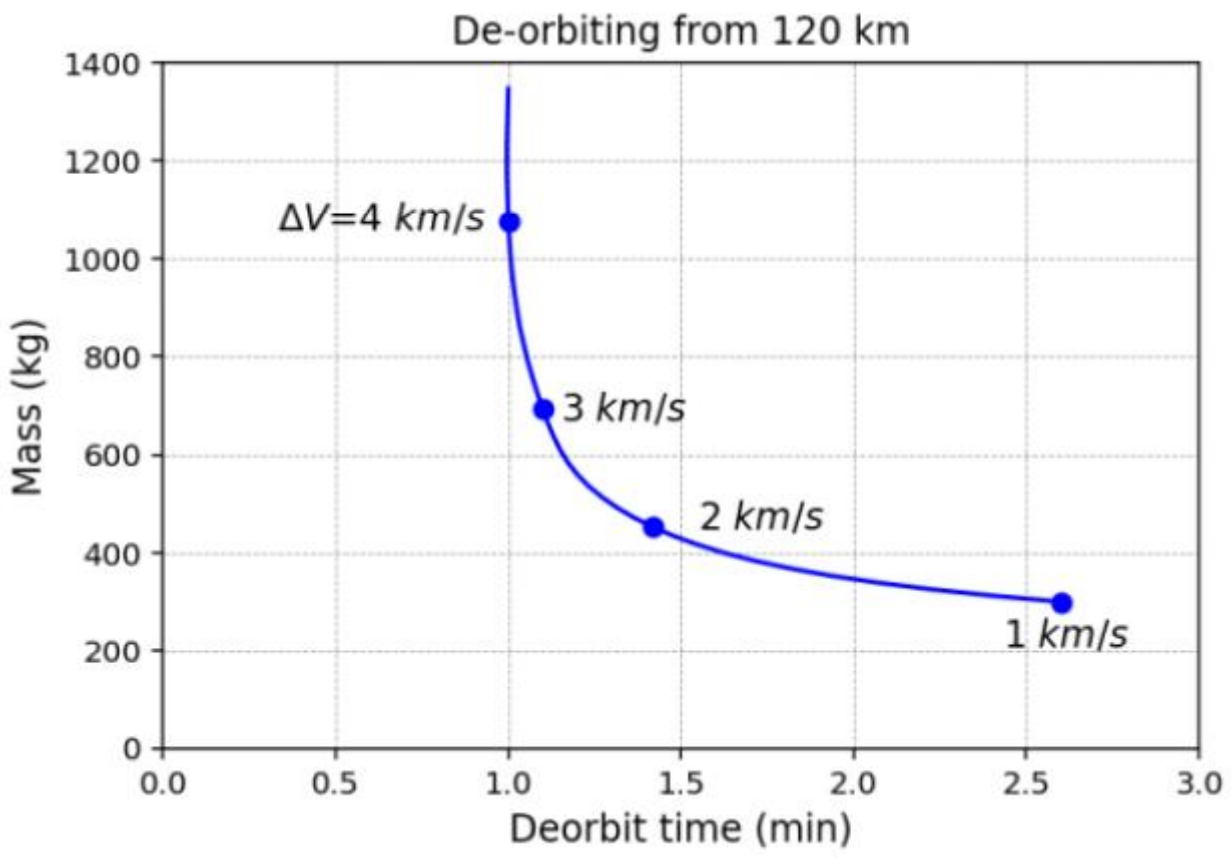

**Figure. 1:** This plot shows how the $\Delta V$ of the deorbit booster affects the total mass of the FOBS weapon and its deorbit time to a point on the ground track of the orbit, assuming the weapon is orbiting at 120 km and the mass of the RV is 200 kg. In this case, the full $\Delta V$ is used to create a downward rather than lateral velocity. The deorbit time shown starts at ignition of the booster and ends with impact. The assumed exhaust velocity of the booster is $V_e$ = 2.75 km/s, its average acceleration is 8g's, and the ballistic coefficient of the RV is $\beta$ =10,000 km/m$^2$. The total FOBS weapon mass scales with the RV mass.

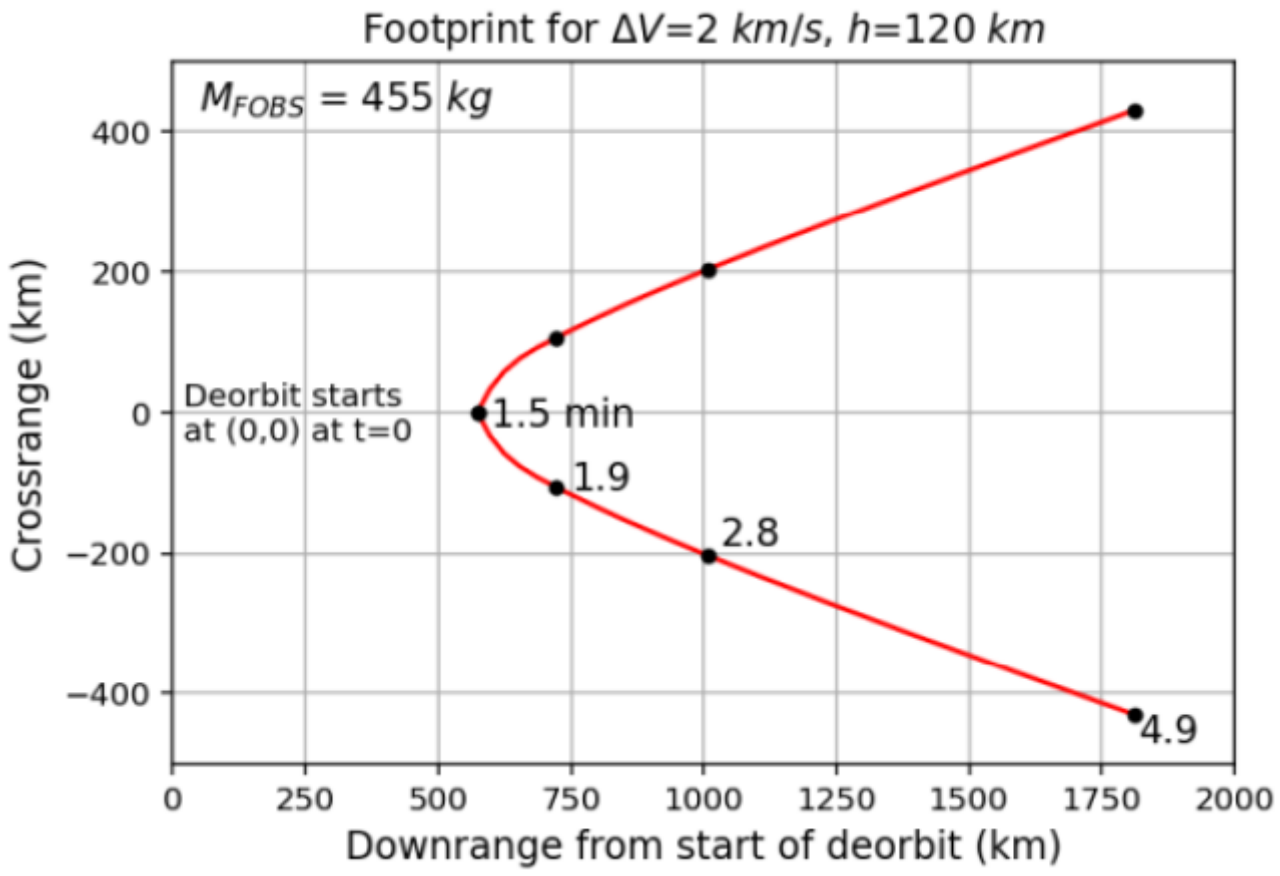

**Figure 2**: This plot shows the impact points of a FOBS weapon orbiting at 120 km with a deorbit booster having a total $\Delta V$ = 2 km/s. As part of the $\Delta V$ is used to give the RV a lateral velocity, impact occurs at some crossrange distance from the ground track of the orbit (which is the x-axis in the plot, with the FOBS weapon moving to the right). The booster ignites at (0,0) at time $t$ = 0. The curve shows the possible ground impact locations, with the points labeled by their deorbit time in minutes. The point labeled "1.5 min" results from the full $\Delta V$ = 2 km/s being used to deorbit, so that the crossrange distance of impact is zero. The points for 1.9, 2.8, and 4.9 min correspond to 0.5, 1, and 1.5 km/s of the total $\Delta V$ being used to give crossrange distances, respectively. $\Delta V$ = 2 km/s gives a FOBS weapon mass of 455 kg, assuming an RV mass of 200 kg. Other details as in Figure 1.

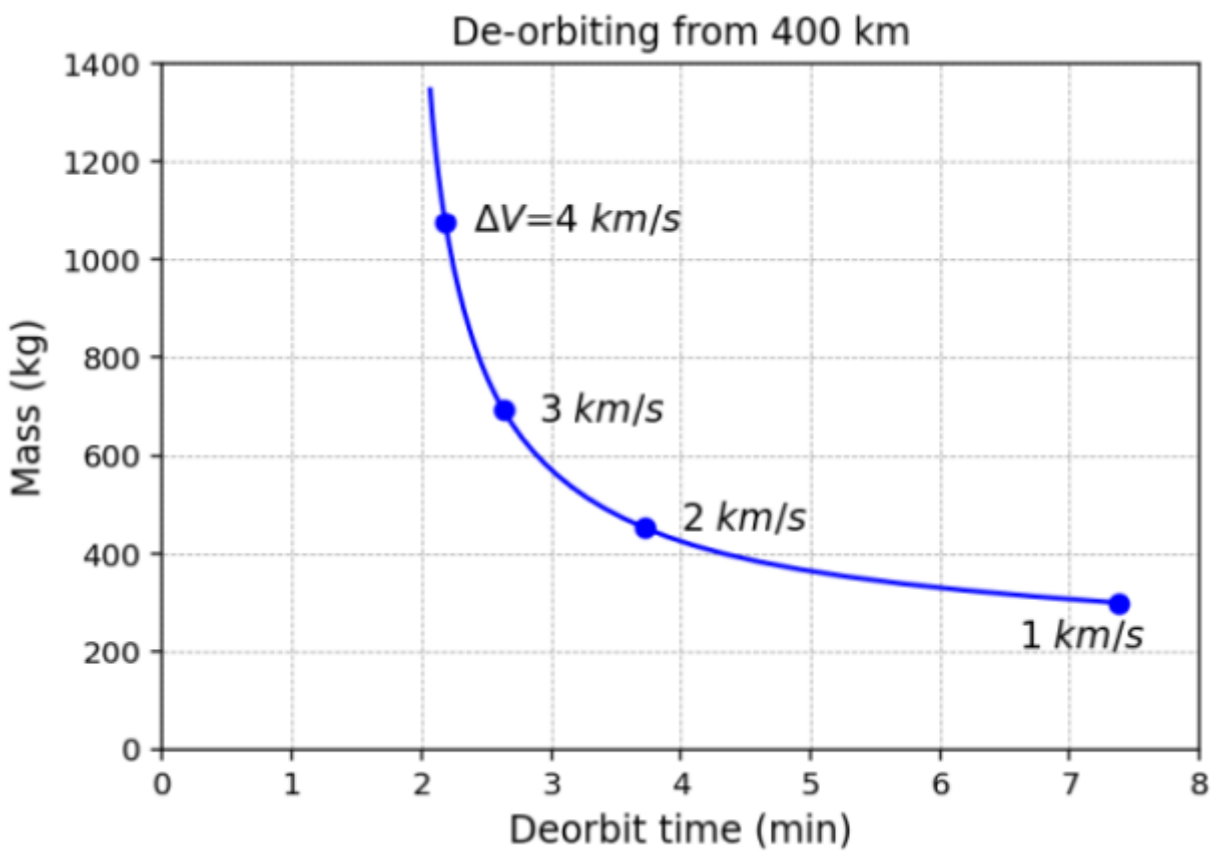

**Figure 3:** Similar to Figure 1, but for an orbital altitude of 400 km.

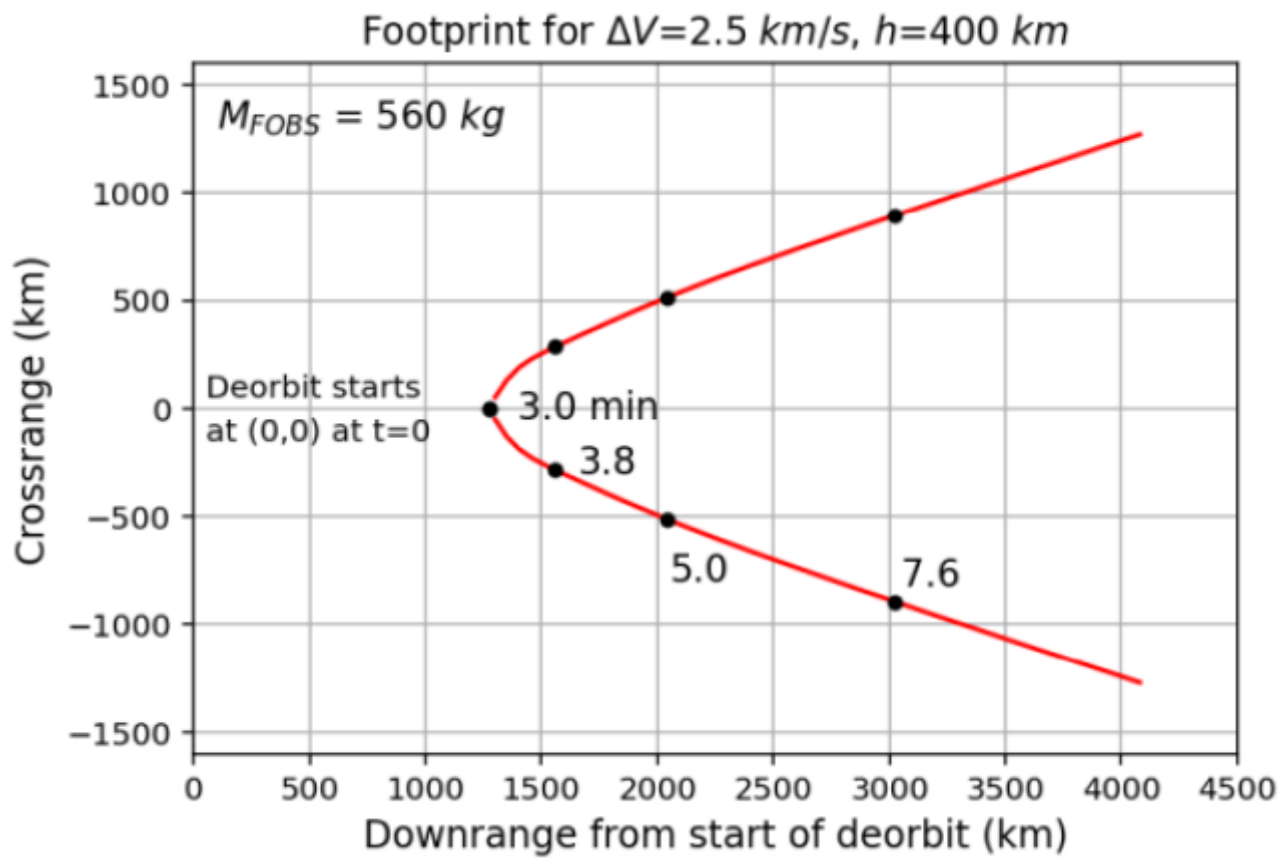

**Figure 4**: Similar to Figure 2, but for a deorbit booster with $\Delta V$ = 2.5 km/s and an orbital altitude of 400 km. The FOBS weapon has a mass of 560 kg, assuming a 200 kg RV mass. The point labeled “3.0 min” results from the full $\Delta V$ = 2.5 km/s being used to deorbit; the points for 3.8, 5.0, and 7.6 min correspond to 0.5, 1, and 1.5 km/s of the total $\Delta V$ being used to give non-zero crossrange distances, respectively.

### 3.3 Adding lift during reentry

We consider two ways of using lift forces during reentry. In each case, we assume the RV can generate lift forces described by a non-zero value of lift-to-drag (L/D).

For the first case we assume the RV follows a typical reentry trajectory but uses lift forces to maneuver as it descends. This maneuvering could be used to increase accuracy, to move laterally to a target, to increase the steepness of the descent to reduce deorbit time, and/or to evade interceptors. The RV can generate lift either by creating an angle-of-attack (the angle between its body axis and its velocity) to produce asymmetric atmospheric forces, or by using wings or other lift surfaces that can be oriented to produce atmospheric forces in a lateral direction.

We analyze this case by integrating the equations of motion in the Appendix with a non-zero L/D term. We find that in the cases considered here, the time the RV travels in regions of significant atmospheric density is short enough that the extent of lateral maneuver is relatively small. In the cases considered here, we estimate crossrange displacements of less than 40 km, assuming L/D = 2. Lift can be used in this way to maneuver near a target, but not to significantly increase the crossrange that is achievable with the deorbit booster. The calculations below assume the RV does not use lift to maneuver laterally during reentry.

### 3.4 Releasing a hypersonic glide vehicle

The second way to add lift during reentry is to put a vehicle with high L/D on a trajectory appropriate to a hypersonic glide vehicle as it reaches altitudes of about 50 km. China is reported to have done this during an August 2021 test, although there appear to be questions about whether the reentering vehicle was a true hypersonic glide vehicle.[23]

Releasing a typical hypersonic glide vehicle would require executing a pull-up maneuver during reentry when the vehicle reaches about 50 km altitude to rotate its velocity and put the vehicle on a horizontal trajectory.[24] The pull-up maneuver would reduce the RVs speed by more than a kilometer per second, so horizontal glide of a FOBS vehicle would likely start with a speed of about 6.5 km/s or less.

A vehicle with this speed could reach long distances both in range and crossrange. For example, consider a glider with L/D = 2 and an initial speed of 6.5 km/s following pull-up, and that dives to the ground when its speed falls to 3 km/s (see below). This vehicle could reach a range of more than 6,000 km and maximum crossrange of about 1,400 km.[25]

Delivering hypersonic vehicles in this way, however, does not make sense for several reasons:

- Performing a pull-up maneuver and gliding to long distances gives long flight times, so they would not be useful for short-time-of-flight attacks. The flight time of the above example flying to its full range would be more than 20 minutes. If there was a reason to use glide vehicles there is therefore no compelling reason to launch them as FOBS. In most cases, the attacker would do better to launch RVs on missiles heading toward the targets and not rely on gliding.

- Designing gliders able to withstand long sustained heating at these high speeds would increase their mass relative to an RV on a typical reentry trajectory. That additional mass is a bigger problem for FOBS delivery because the launcher must place that mass in orbit, rather than on a ballistic trajectory. Moreover, we note that test failures apparently caused by the very high heating at these speeds over long flight times led the US to cancel its program to develop long-range hypersonic weapons, and to refocus on shorter range systems.[26]

- Gliders, like FOBS, could underfly current long-range defenses. Reentry analysis, however, shows that for the glider to also be able to evade terminal defenses as it dives on a target, it must begin its dive by the time its speed has dropped to about 3 km/s.[27] That limits the range that a glider can usefully reach.

For these reasons we do not consider the case of a FOBS weapon releasing a hypersonic glide vehicle.

## 4. Trajectory Calculations

In this section we give examples of the three types of FOBS attacks discussed above and estimate the delivery and warning time in each case. These include a FOBS attack using multiple ballistic missiles launching weapons on 120-km altitude orbits to underfly current strategic missile defenses, and two types of short-warning attacks disguised as satellite launches.

### 4.1 FOBS Attacks to Avoid Missile Defenses

These launches do not attempt to be covert, so multiple FOBS could be launched at the same time toward a set of targets using ballistic missiles. The attacker would instead launch these weapons into 120 km altitude orbits, which is too low for current strategic defenses to intercept even if the FOBS weapons can be seen and tracked.

Figures 5 and 6 assume that a target at a specified range is attacked by warheads launched by a ballistic missile, either on an MET or as a FOBS orbiting at 120 km, and compare both the payload that could be delivered by these methods to that range and the time of delivery from launch to impact. For specificity, these calculations assume the ballistic missile is a Minuteman III (MMIII) ICBM that can launch a payload of 900 kg to a range of 10,000 km on an MET. We then vary the range to the target and determine what payload this missile could launch on an MET to this new range, and the flight time of that attack. The FOBS is assumed to be launched in a direction that places the target on the ground track of the FOBS orbit, so that deorbiting does not require crossrange maneuvering.

Figure 5 shows the payloads and flight times of the MET and FOBS as a function of range, and Figure 6 shows the ratio of the two curves.

A 10,000 km range MET for the MMIII has a burnout speed of 6.90 km/s. Launching a FOBS weapon into a circular orbit at 120 km requires accelerating the payload to 7.84 km/s, which an MMIII can do with a payload of about 510 kg. Once in orbit, the weapons can fly to any range, so the payload will be the same for all ranges. The times for launching the FOBS weapon into orbit and deorbiting it to hit the target will be independent of the total range, so the flight time will vary with the length of time it is in orbit.

The effective payload reduction will be even greater for the FOBS than shown in Figure 6 because the payload must also include the mass of the deorbit booster, which therefore reduces the mass of RVs the missile can carry. A small booster mass will give a small $\Delta V$, which will give a deorbit trajectory that is less steep and has a longer path within the atmosphere. This increases deorbit time and reduces the RV's speed when it reaches low altitudes. If the speed is too low, the RV may be vulnerable to interception by endo-atmospheric interceptors like current PAC-3 or SM-6 systems. To avoid this, the attacker would want to keep the RV speed above about 2 km/s when the RV reaches an altitude of about 5 km.[28]

For the case we consider, with the booster as described above and $\beta = 10{,}000$ kg/m$^2$ for the RV, this requires the booster to have $\Delta V$ of about 1.5 km/s if the RV uses no lift during reentry. This value can be reduced to $\Delta V = 0.5$ km/s or less if the RV has $L/D = 1$ that it uses to increase the steepness of the reentry orbit within the atmosphere. Assuming an RV mass of 200 kg, $\Delta V = 1.5$ km/s gives a FOBS weapon mass of 370 kg, while $\Delta V = 0.5$ km/s gives a FOBS weapon mass of 250 kg.

Combining the payload reduction from 900 kg for a 10,000 km MET to 510 kg for a FOBS to reach orbital trajectory, with the fraction of payload that is used for the deorbit booster, the effective payload for carrying RVs is reduced to less than half compared to the MET for the case of $\Delta V = 0.5$ km/s, and less than a third for $\Delta V = 1.5$ km/s. In this case, using lift during reentry to reduce the FOBS weapon mass as described in the previous paragraph can have an important consequence: A MMIII that can place a payload of 510 kg into orbit can launch two FOBS weapons with $\Delta V = 0.5$ km/s but only a single FOBS weapon with $\Delta V = 1.5$ km/s.

Using FOBS to evade missile defenses in this way can lead to a significant payload penalty per missile compared to an MET, although it would still allow a missile to deliver at least one nuclear warhead to a target at any range. Using FOBS would give a shorter flight time than an MET, although likely not short enough to allow a decapitation attack, as discussed above.

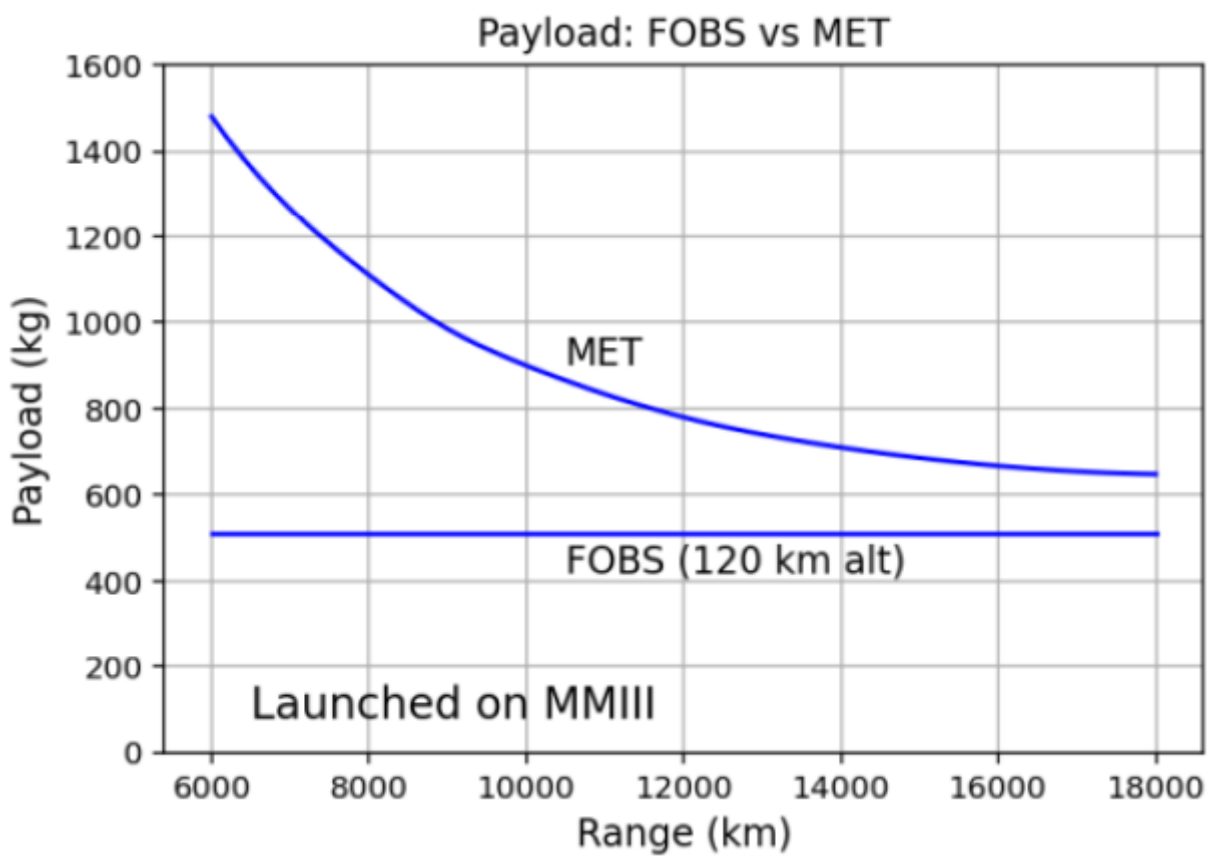

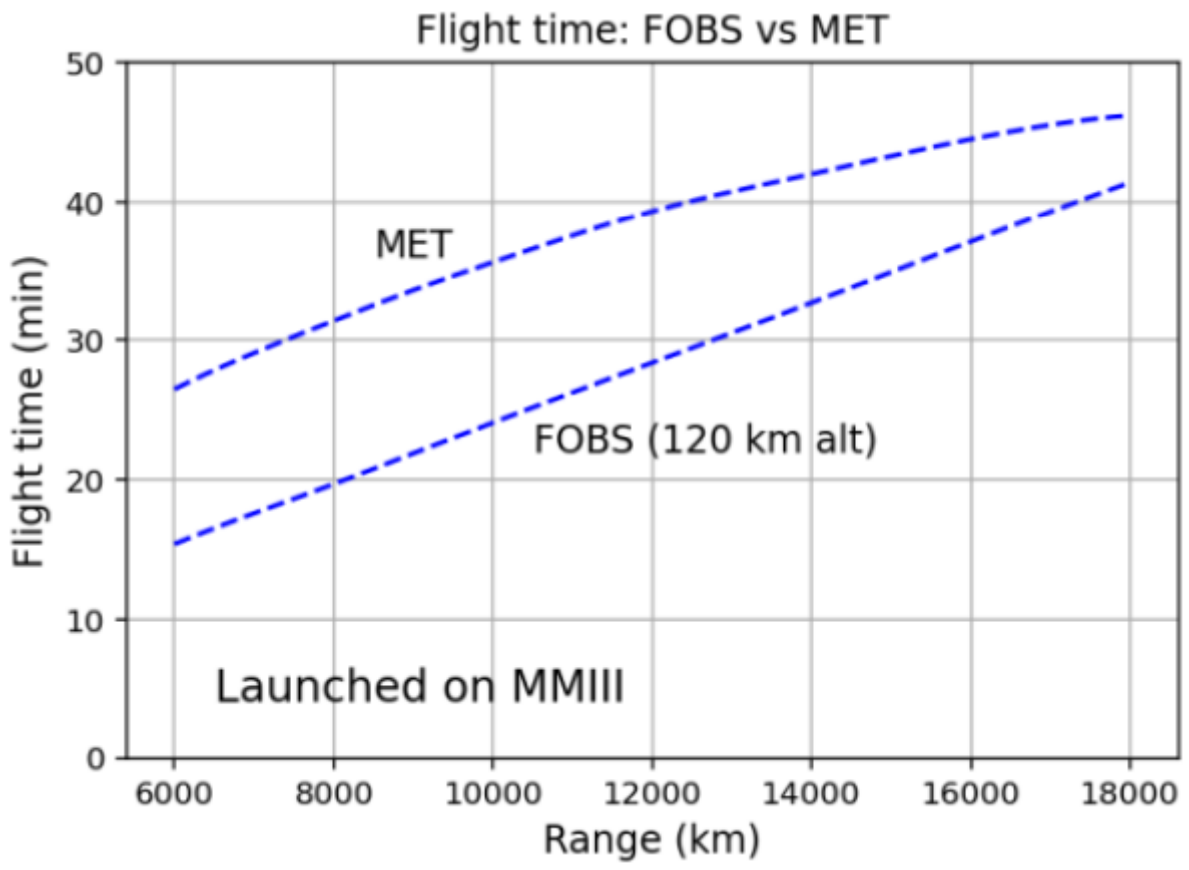

**Figure 5.** These plots show the payload (left) and flight time (right) of an attack using ballistic missile on an MET or a FOBS, in both cases assuming the missile is similar to a Minuteman III. The FOBS weapon is placed in an orbit at 120 km altitude. The MET payload can be increased as the range decreases since the required burnout speed of the missile is smaller for shorter ranges.

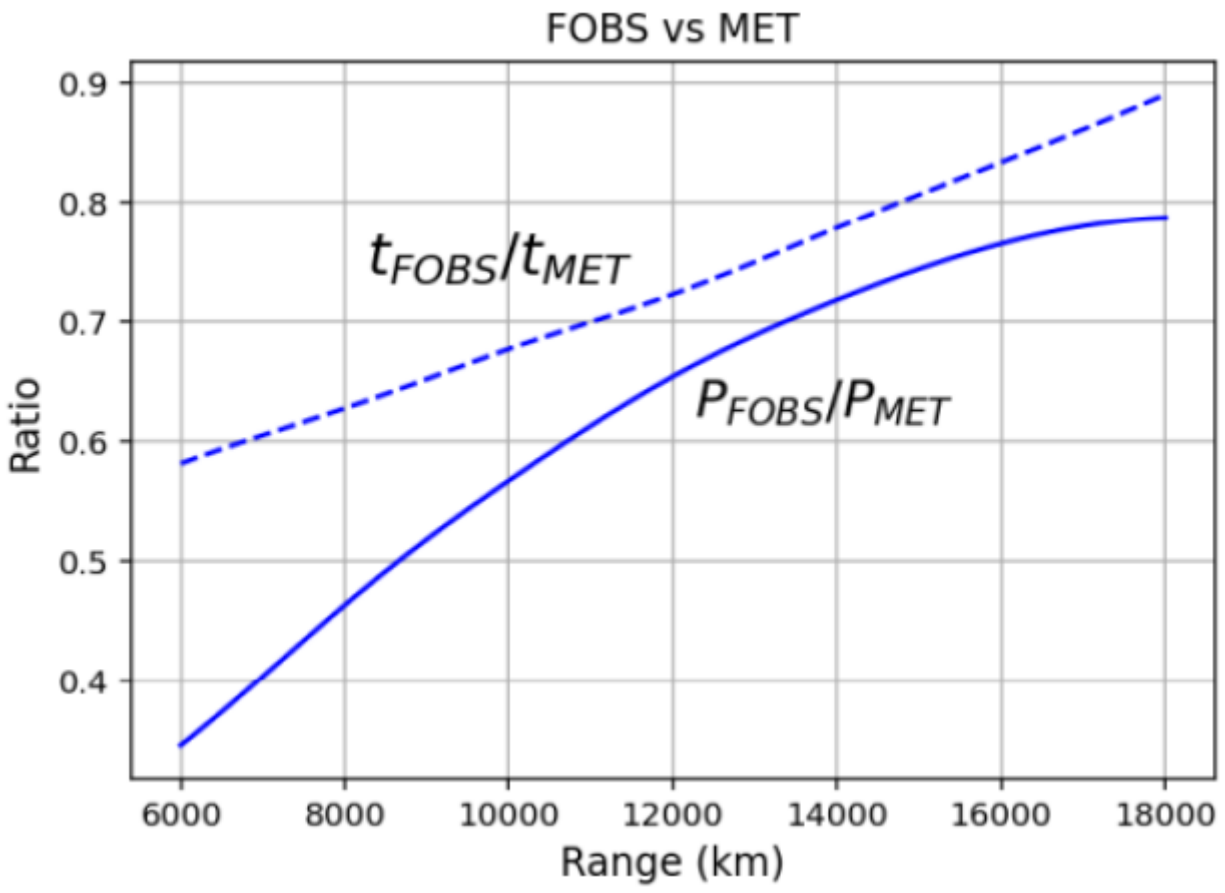

**Figure 6.** This plot shows the ratio of the values in Figure 5. The solid line shows the ratio of the FOBS weapon mass to the MET payload mass for each range, and the dotted line shows the ratio of the flight times of the FOBS and MET over each range. The effective payload for FOBS warheads will be further reduced by the fraction that is required for the FOBS deorbit booster.

## 4.2 Very-short-warning Attacks Disguised as Satellite Launches

We next consider two types of attacks that are disguised as satellite launches, intended to give very short warning of attacks on a small number of key targets, as discussed above. In both cases the adversary is assumed to launch FOBS on SLVs into orbits it has used for past satellite launches, so they appear to be standard satellite launches. The orbits are chosen with inclinations that will carry the FOBS weapons close to the location of the targets.

A list of satellite launches and details about them can be found in Jonathan McDowell's databases.[29] This includes the launch site, launch vehicle, type of payload, and inclination and altitude of the orbit.

US space-based satellites sensors would detect the launch of the SLV, but if this is interpreted as a satellite launch, the earliest warning of an attack would come when the firing of the deorbit booster is detected. If the thruster burn is interpreted as an orbit maneuver, the US might not have clear warning until tracking begins to establish the deorbit trajectory and estimated target. The times given in the analysis below represent the maximum warning time, which is the time between the beginning of the deorbit burn and impact at a target.

To illustrate this approach, we analyze hypothetical launches by Russia and China against targets near Washington, DC, and Omaha, NE. For each case we pick a launch site and orbit of a past satellite launch, choosing an orbit that passes near the targets, and assume the SLV launches one or more FOBS weapons into an orbit with the same inclination and altitude. The FOBS boosters then deorbit the warheads toward the targets.[30] The calculations assume the RV does not use lift to maneuver laterally during reentry, as discussed above.

The first type of attack we consider assumes a single SLV launch into an orbit that carries the FOBS weapon close enough to both Washington and Omaha that lateral thrust from the deorbit booster can cause RVs to impact at both targets in less than 10 minutes. Using a single SLV limits the number of FOBS weapons that can be launched in the attack. Moreover, adding significant lateral maneuvering to the weapons increases the mass of the deorbit boosters, which further reduces the number of warheads that can be launched in the attack.

The second type of attack assumes that two SLVs are launched several hours apart from different sites, so that they are again interpreted as satellite launches. The orbits and times of launch are chosen so that at some time within 24 hours after the first launch each orbit will pass over the targets at the same time, so that they can be deorbited to give simultaneous attacks. By choosing orbits that pass over the targets, this method of attack reduces the need for lateral maneuvering and therefore the mass of the FOBS weapon. The reduction in mass and the use of two SLVs increases the number of warheads that can be used in the attack. It does, however, increase the time after launch before the attack on the targets can be carried out.

#### 4.2.1 Single SLV launch with warheads making fractional orbits

Here we consider a hypothetical attack on Washington and Omaha using launches from the Vostochny space launch center in Russia and the Jiuquan space launch center in China.

For each launch site, we identify a past satellite launch into a low-altitude orbit that passes near the two targets on its first orbit. The data from early warning satellites allows the United States to determine the type of SLV being launched, so the adversary would likely want to use the same launch vehicle it used with the satellite launch it is mimicking.

We then choose a point on the orbit where the FOBS weapon would fire its deorbit booster so that a point of its footprint falls on one of the targets, and calculate the deorbit time. We then repeat this process so that the footprint of a second FOBS weapon from the same orbit falls on the second target. In general, reaching different targets will require different values of $\Delta V_{tot}$ for the two weapons and different location to start the deorbits. In each case the goal is to find a combination of short deorbit time and low FOBS weapon mass, because the mass determines how many weapons the SLV can launch. Estimating the payload the assumed SLV can launch into the assumed orbit then determines what combination of FOBS weapons it can carry.

#### 4.2.1.1 Launch from Vostochny, Russia

As an illustration, from the list of satellite launches from Russian launch sites into near-polar orbit in recent years we chose a launch from 2022 with the COSPAR number 2022-139B. A Soyuz-2 1B/Fregat launched the GONETS-M 34L civil communications satellite north from the Vostochny launch site into a 200 x 200 km initial orbit with inclination of 82.5°.[31] The orbit is shown in Figure 7 at the time it passes near Washington.

At 34 minutes after launch the Fregat stage burned to put the satellite into a 200 x 1500 km orbit, and after 98 minutes it burned again to put it into its final 1482 x 1500 km orbit.[32] A FOBS weapon launched into a similar orbit would have an altitude of 200 km when it passed over Washington on its first orbit.

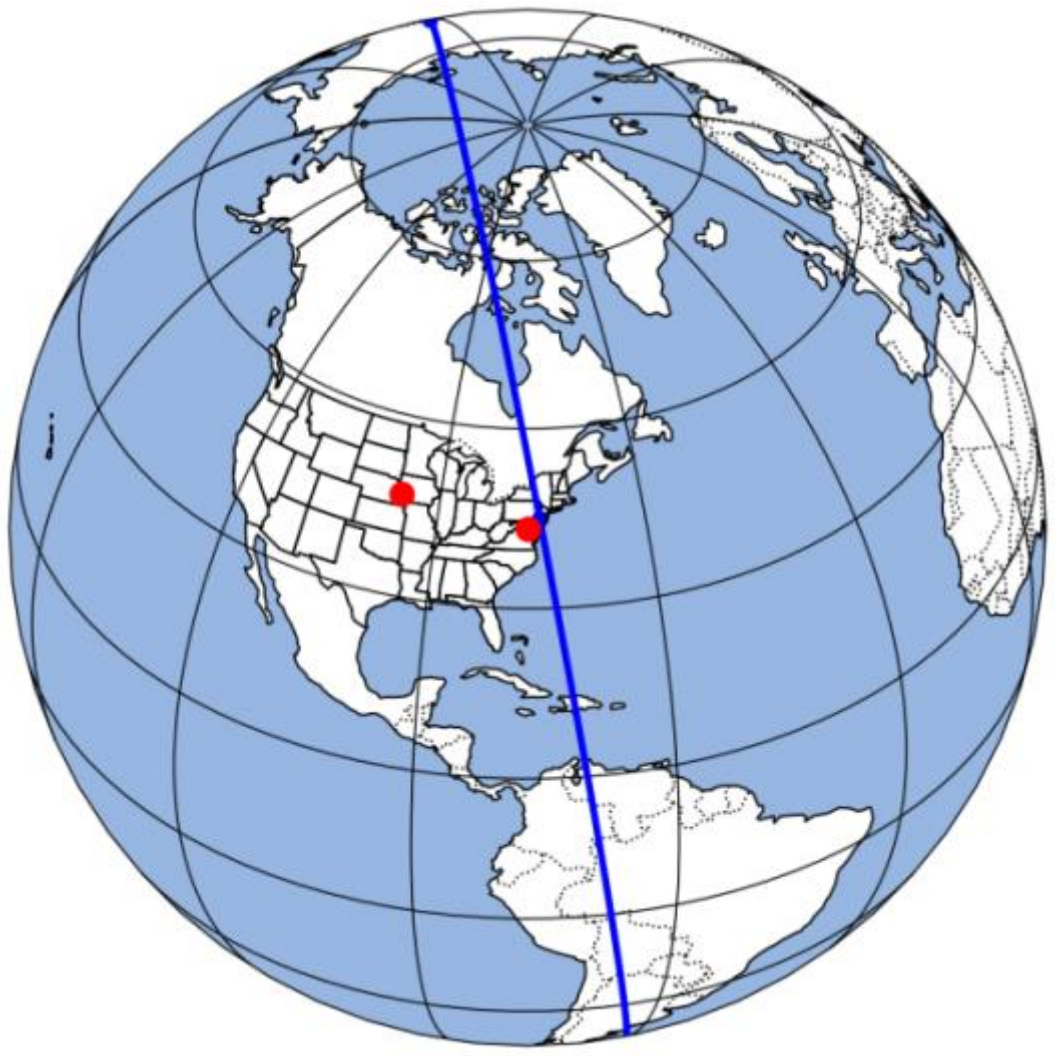

**Figure 7**. The line shows the ground track of a FOBS weapon launched from Vostochny on an orbit like 2022-139B, and is shown at the time the weapon passes near Washington on its first orbit, about 30 minutes after launch. The FOBS would be traveling north to south. The orbit has an inclination of 82.5° and an altitude of 200 km on its first pass. The dots show the locations of Washington and Omaha.

Attacking Omaha would require the FOBS booster to give the RV significant lateral speed. Figure 8 assumes a deorbit booster with $\Delta V_{tot}$ = 4 km/s, which fires when the FOBS weapon is over the dot shown in northern Canada (72.46°, -93.57°). The footprint curve shows the possible impact points, which pass through Omaha. The time from the beginning of the deorbit burn to impact at Omaha is 8.5 minutes. The FOBS weapon mass in this case is 1,075 kg assuming a 200 kg RV. Reducing $\Delta V_{tot}$ to reduce the vehicle mass will increase the time to impact.

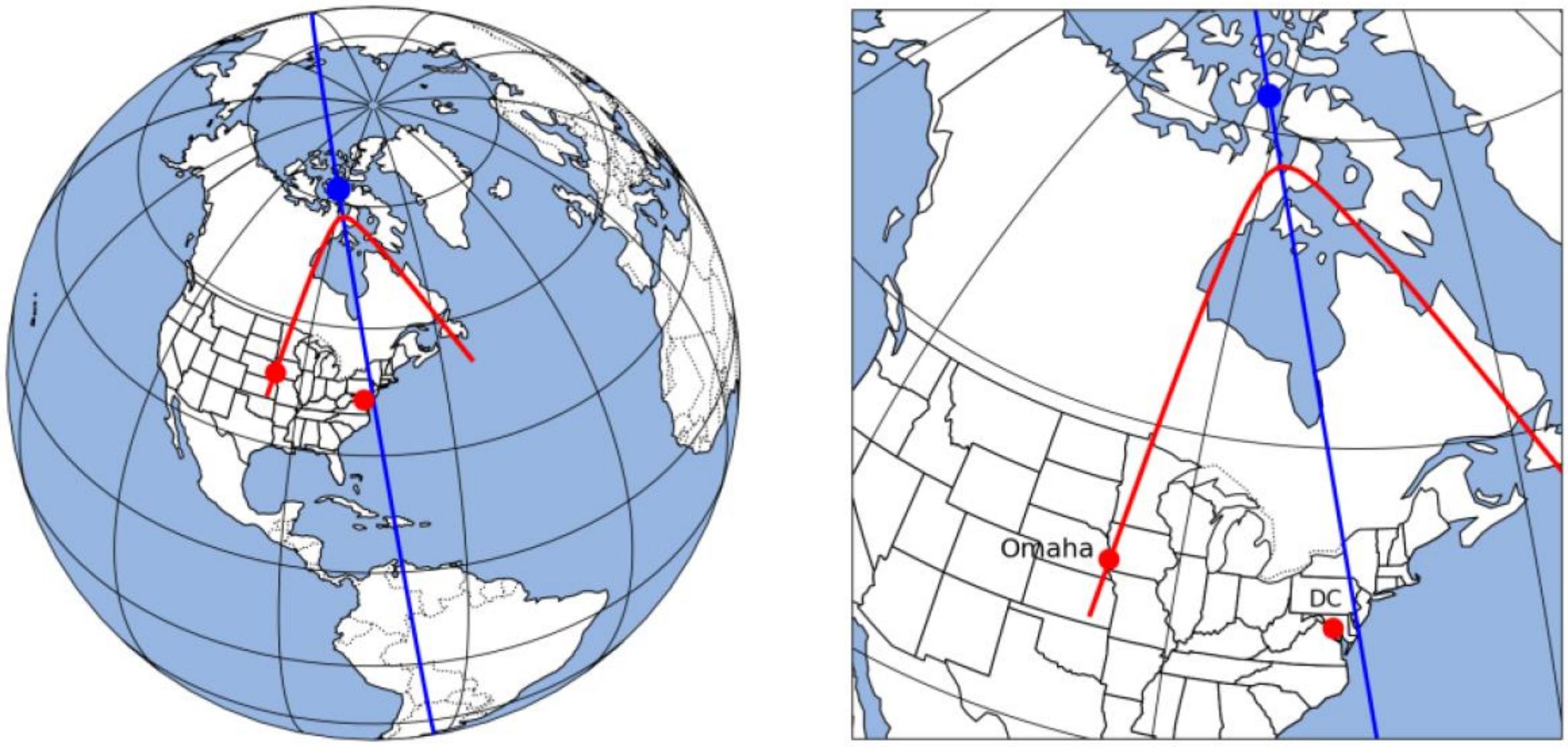

**Figure 8:** The north-south curve is the ground track of the orbit shown in Figure 7, launched from Vostochny, Russia. The dots are Washington and Omaha. The dot over northern Canada is where the deorbit booster fires to target Omaha, about 21 minutes after launch. The booster provides $\Delta V_{tot}$ = 4.0 km/s, and the v-shaped curve is the possible impact points of the RV. In this case, impact will occur at Omaha 8.5 minutes after deorbit begins. We refer to this trajectory as VOS-to-Omaha in Figure 14 and Table 3.

Attacking targets in or near Washington with a FOBS weapon on this orbit would require very little lateral maneuvering, so essentially all the $\Delta V$ of the FOBS booster could be used for deorbiting. A booster with $\Delta V$ = 1.5 km/s would give a deorbit time of 2.7 minutes from this altitude, and result in a FOBS weapon mass of 370 kg assuming a 200 kg RV. The deorbit burn for this FOBS weapon would start at a point 1,080 km along the orbit north of Washington, which is a ground range of 2,850 from the deorbit point for attacking Omaha, and it would take 6.1 minutes for the weapon to travel that distance in orbit. Impact at Washington would therefore occur 8.8 minutes after deorbit begins for Omaha.[33]

The Soyuz-2-1B SLV has a reported payload capacity to sun-synchronous orbit of about 6,900 kg.[34] It could therefore launch three FOBS weapons toward Omaha and nine or 10 toward Washington—more than sufficient to deliver two warheads against each of the five targets in the decapitation outline outlined above.

#### 4.2.1.2 Launch from Jiuquan, China

We consider a similar hypothetical launch from Jiuquan, China, in this case the 2026 launch of a Bai Xiang test satellite on a Lijian-2 launcher. This satellite, 2026-066A, was launched south from Jiuquan into a 217 x 604 km orbit with inclination of 85.0$^{o}$ and period of 92.6 minutes. A series of burns starting about 23 hours after launch placed it into a 584 x 621 km orbit.[35] Figure 9 shows the orbit at the time it passes between Omaha and Washington on its first orbit; the satellite is traveling south to north across North America. The perigee of this orbit occurred at 19.18$^{o}$ N, 73.98$^{o}$ E over northern India, and a FOBS weapon on this orbit would have an altitude of about 475 km and a speed 7.59 km/s as it passed near Omaha on its first orbit.

The left panel of Figure 10 shows a FOBS weapon attacking Washington from this orbit. It assumes a deorbit booster with $\Delta V_{tot}$ = 3.0 km/s, which fires over the dot in the bottom left of the figure (13.63$^{o}$, -92.49$^{o}$). The footprint shows the possible impact points, which pass through Washington. The time from the beginning of the deorbit burn to impact on Washington is 7.7 minutes. The FOBS weapon mass in this case is 690 kg, assuming a 200 kg RV.

The right panel of Figure 10 shows a FOBS weapon deorbiting to attack Omaha from this orbit. In this case, the deorbit booster again has $\Delta V_{tot}$ = 3.0 km/s and fires over the dot in the Gulf of Mexico, which is 1,100 km north of the deorbit point for Washington. The footprint shows the possible impact points of this FOBS. The time from the beginning of the deorbit burn to impact at Omaha is 4.7 minutes, but deorbit begins 2.6 minutes after the deorbit toward Washington, so impact will occur 7.3 minutes after first warning of an attack.

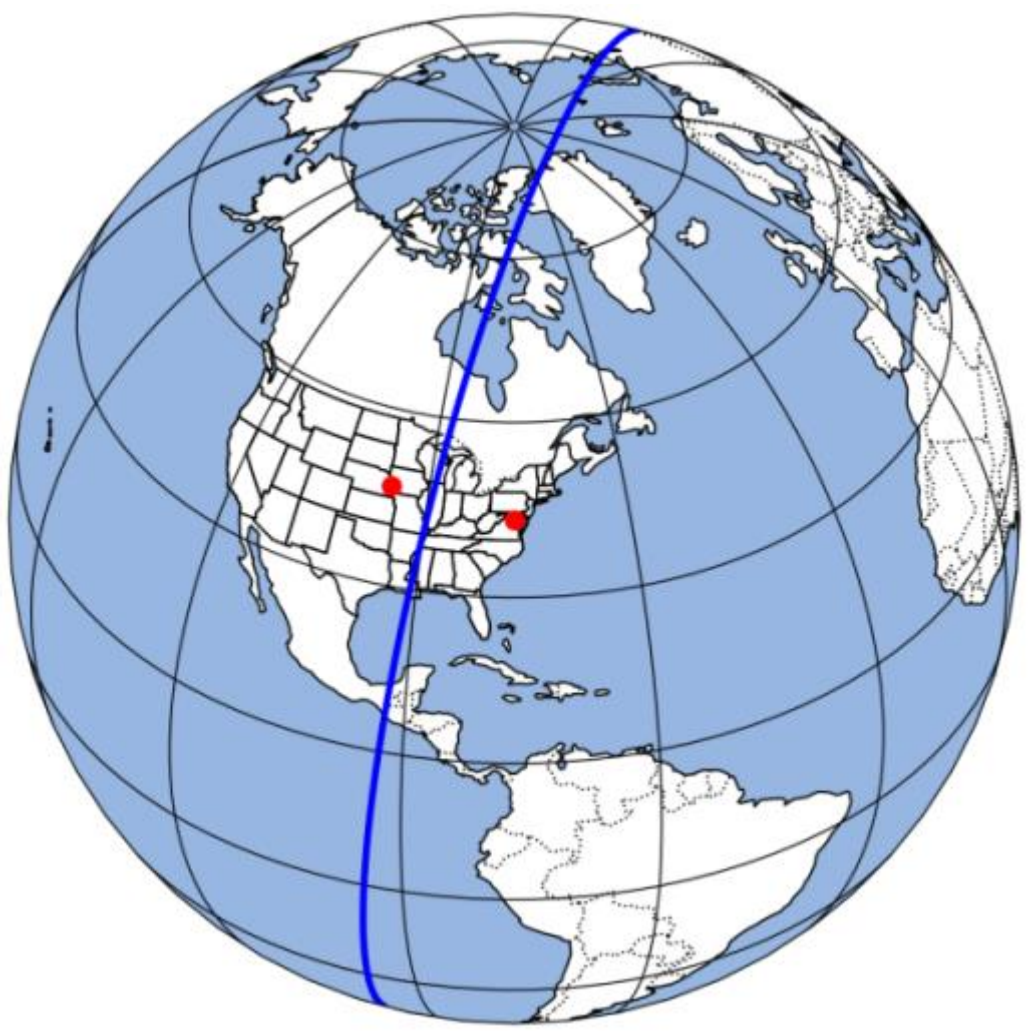

**Figure 9**. The curve is the ground track of a FOBS launched south from Jiuquan, and is shown at the time the FOBS weapon passes near Omaha on its first orbit, 73 minutes after launch. The FOBS is traveling south to north along the orbit. The dots indicate Washington and Omaha.

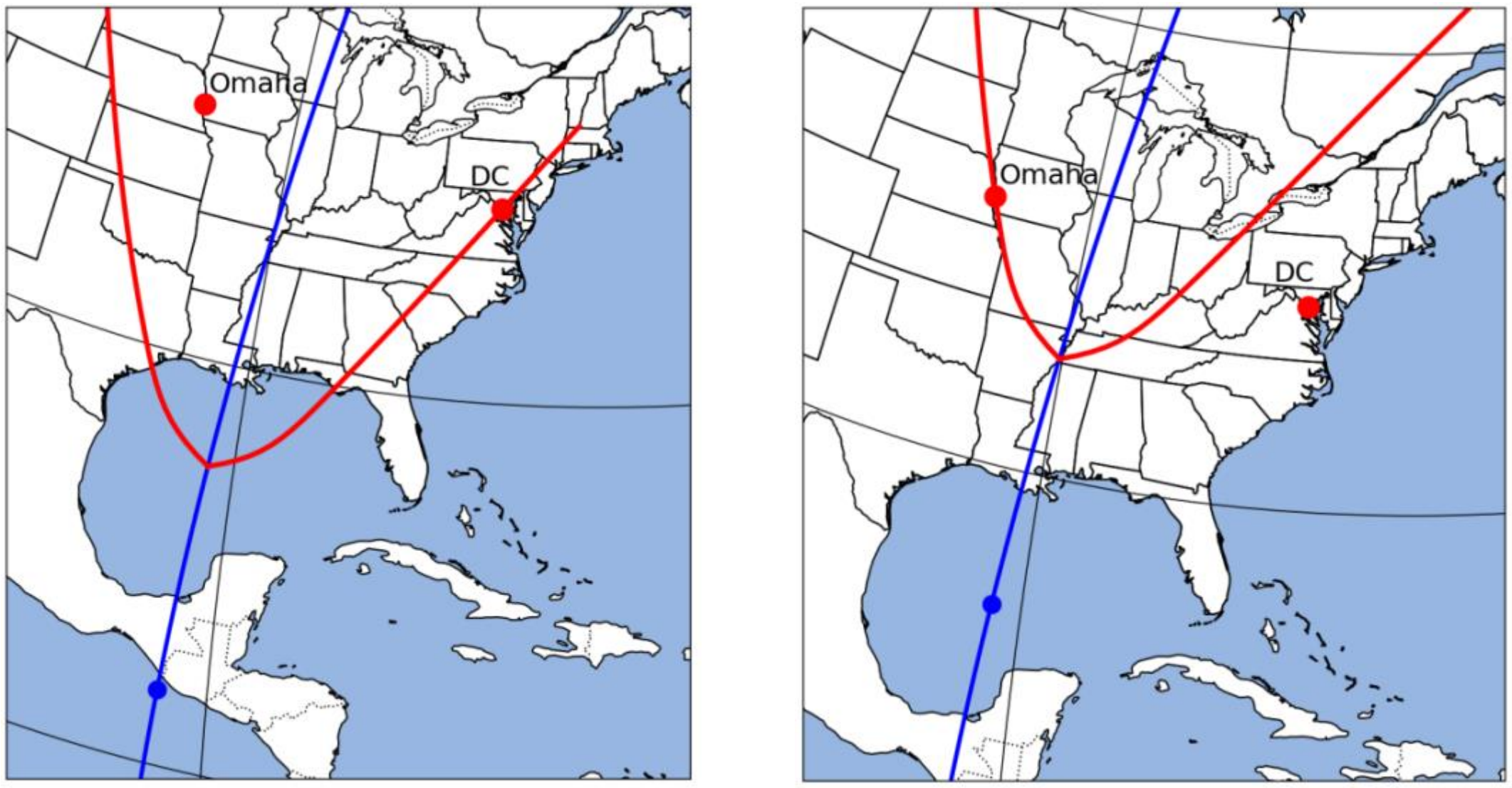

**Figure 10:** These figures show the orbit from Figure 9, which is traveling south to north across the US. The dots are Washington and Omaha. The dot near the bottom of the left panel shows where the deorbit booster fires to target Washington. The booster provides $\Delta V_{tot}$ = 3.0 km/s, and the v-shaped curve is the possible impact points of the RV. Impact on Washington will occur 7.7 minutes after deorbit begins. The right panel shows the FOBS weapon targeting Omaha. It begins deorbiting 2.6 minutes after the Washington FOBS

weapon begins to deorbit, with impact 4.7 minutes later, so impact is also 7.3 minutes after the first warning of attack given by the first deorbit burn. We refer to the left and right trajectories as JQ-to-DC and JQ-to-Omaha, respectively, in Figure 14 and Table 3.

The Lijian-2 SLV can reportedly launch 8,000 kg into a sun-synchronous orbit at 500 km altitude.[36] It could therefore launch 10 to 11 of the FOBS weapons described here—sufficient to deliver two weapons on each of the five targets in the decapitation attack described above.

### 4.2.2 Launches of two SLVs with warheads making multiple obits

A second option for a hypothetical attack on Washington and Omaha assumes that two SLVs carrying FOBS weapons are launched on orbits that pass over the targets after several orbits. These are sometimes called MOBS, or Multiple Orbit Bombardment Systems.[37] To disguise the attack, the attacker is again assumed to choose orbits that have previously been used to launch satellites, and would launch the SLVs from different locations several hours apart, with the launches timed so that the FOBS weapons arrive over the two targets at the same time.

An advantage of this type of attack is that by choosing orbits that pass over the targets, the FOBS weapons will not require much lateral maneuvering so that the deorbit time will be shorter than those above, giving a very short warning time. One disadvantage is that it requires the attacker to launch the first FOBS some hours before the attack takes place, so that the attack cannot be carried out as quickly as a single-FOBS attack. In these examples we consider orbits that allow an attack within about 12 hours of the first launch.

#### 4.2.2.1 Launches from Russia

In this example, we consider a launch from Vostochny targeting Washington and a launch from the Plesetsk space launch center targeting Omaha.

*Launch from Vostochny to Washington:*

The attack on Washington is assumed to use orbit 2022-139B shown in Figure 7 above. As discussed, the FOBS would be launched into a 200 x 200 km initial orbit with inclination of 82.5°. In this case, the FOBS weapon would deorbit midway through its first orbit, about 25 minutes after launch. Assuming a booster with $\Delta V_{tot}$ = 1.5 km/s, deorbit would take only 2.7 minutes. The FOBS weapon would have a mass of 370 kg. The Soyuz-2 1B could carry 18 of these weapons.

*Launch from Plesetsk to Omaha:*

Figure 11 shows orbit 2024-166B from Plesetsk, which used an Angara-1.2 SLV to launch an OO-MKA optical reconnaissance satellite into a 328 x 343 km orbit with inclination 96.8°.[38] Figure 11 shows the ground track as it passes over Omaha on its seventh orbit, 9.6 hours after launch. A FOBS booster with $\Delta V_{tot}$ = 1.5 km/s would give a deorbit time of 4.2 minutes and a weapon mass of 370 kg. The Angara-1.2 can reportedly carry 2,400 kg into a sun-synchronous orbit. It could therefore carry six 370-kg FOBS weapons.[39]

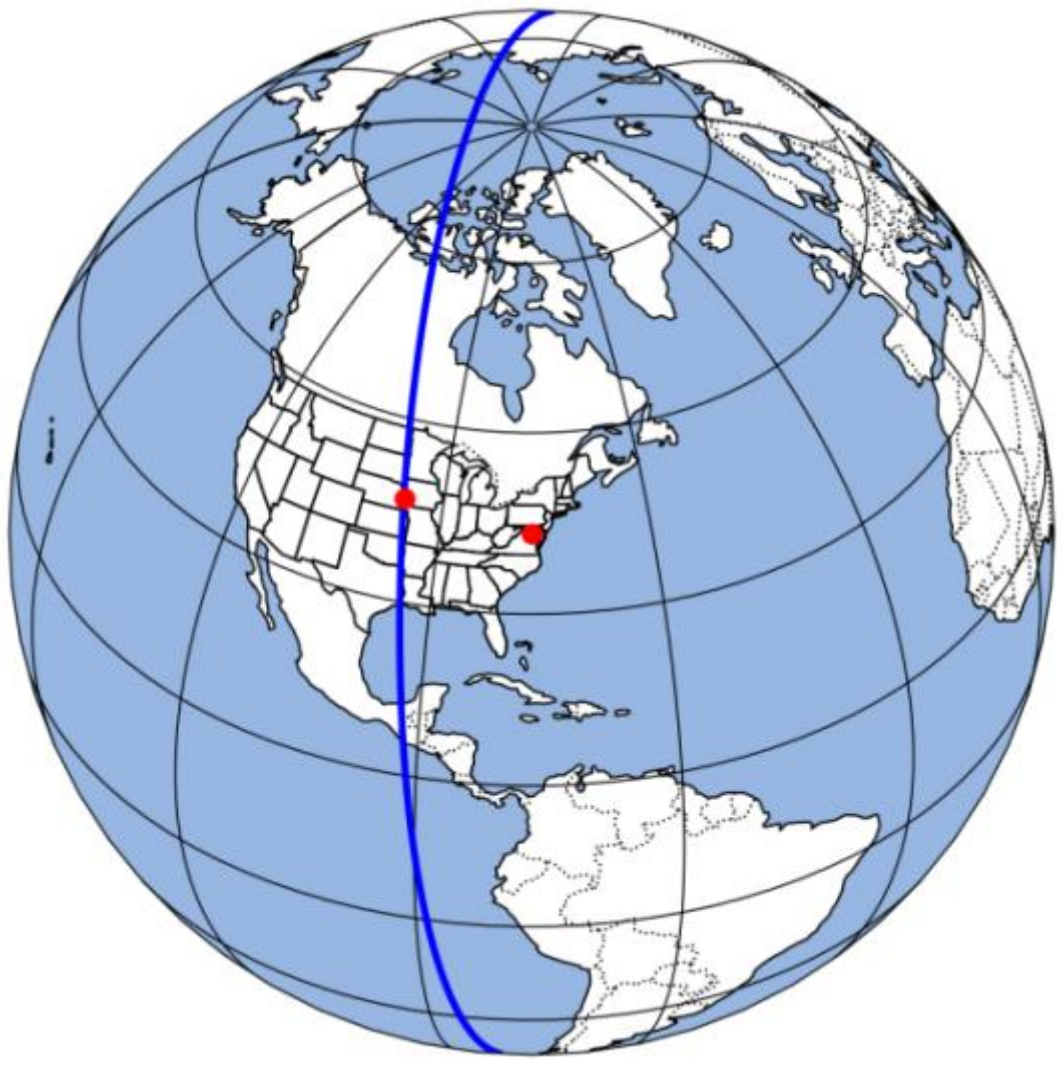

**Figure 11:** This figure shows the orbit of a launch similar to 2024-166B from Plesetsk, Russia, as the FOBS weapon passes over Omaha on its seventh orbit, 9.6 hours after launch.

#### 4.2.2.2 Launches from China

This example considers a hypothetical launch from Jiuquan targeting Washington and from the Xichang space launch center targeting Omaha.

*Launch from Jiuquan to Washington:*

Figure 12 shows the orbit of 2026-128A, which was launched south from Jiuquan and is passing over the US from south to north. In that launch, a Zhuque-2E SLV carried a Qianfan DTC broadband mobile test satellite into a 394 x 405 km orbit at 55.0$^o$ inclination.[40] The orbit passes over Washington on its third orbit, 4.6 hours after launch.

A FOBS booster with $\Delta V_{tot}$ = 1.5 km/s would give a deorbit time of 4.9 minutes and a weapon mass of 370 kg. A booster with $\Delta V_{tot}$ = 2.0 km/s would give a deorbit time of 3.7 minutes and a weapon mass of 460 kg. The Zhuque-2E Block 2 can reportedly carry about 6,000 kg to a 200 km orbit and about 4,000 kg to a 500 kg near-polar orbit.[41] One SLV should therefore be able to launch at least 10 of the FOBS weapons described here.

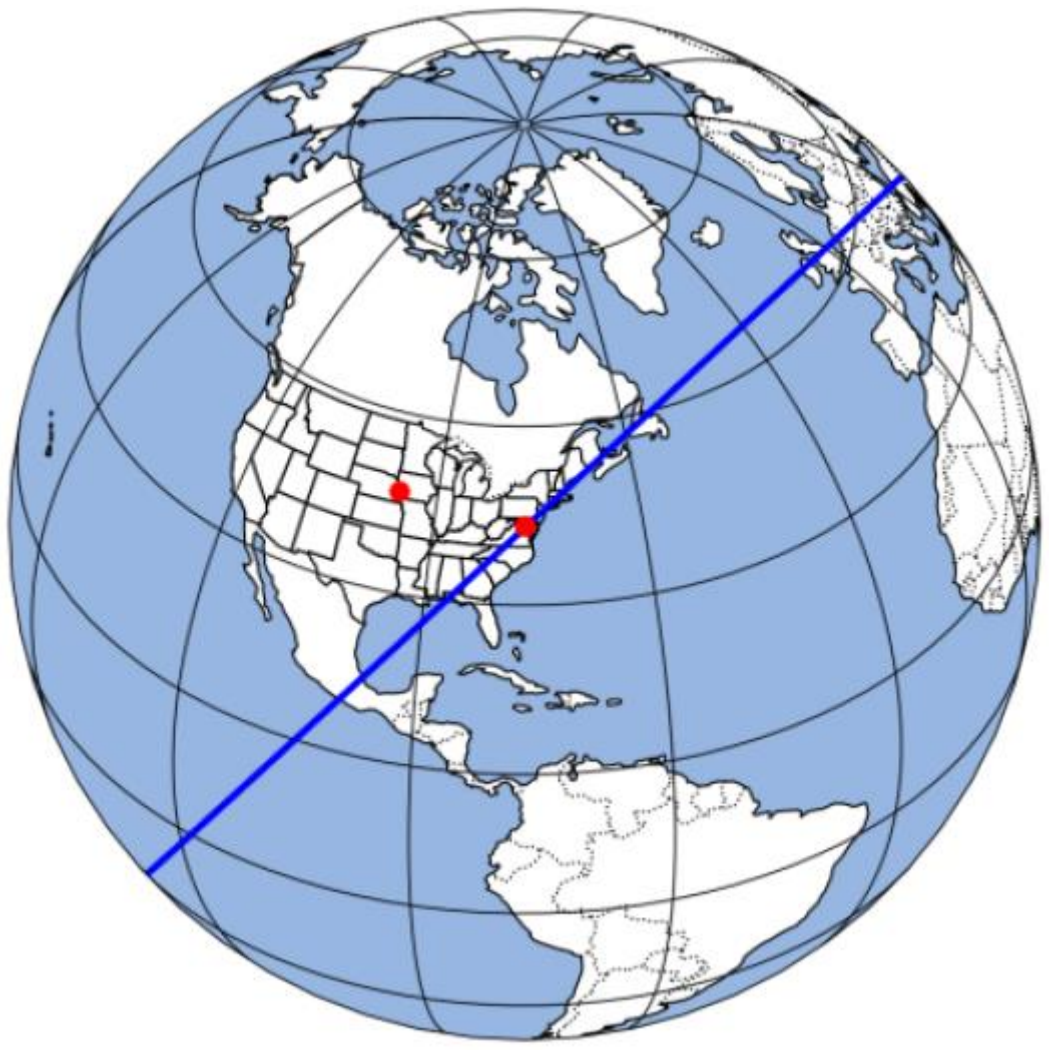

**Figure 12:** This figure shows the orbit of a launch similar to 2026-128A from Jiuquan, China, as the FOBS weapon passes over Washington on its third orbit, 4.6 hours after launch.

*Launch from Xichang to Omaha:*

Figure 13 shows orbit 2026-120A, which was launched south from Xichang using a Long March 2D SLV to orbit a Xingwang Jishu SW-10, which is a test satellite for a broadband constellation. The orbit is 440 x 453 km at 55.0º inclination.[42] The orbit passes over Omaha on its eighth orbit, 12.4 hours after launch.

A FOBS booster with $\Delta V_{tot}$ = 1.5 km/s would give a deorbit time of 5.4 minutes and a weapon mass of 370 kg. A booster with $\Delta V_{tot}$ = 2.0 km/s would give a deorbit time of 4.1 minutes and a weapon mass of 460 kg. The Long March 2D can launch about 3,000 kg into this orbit, so it could carry six to eight of these FOBS weapons, depending on their mass.[43]

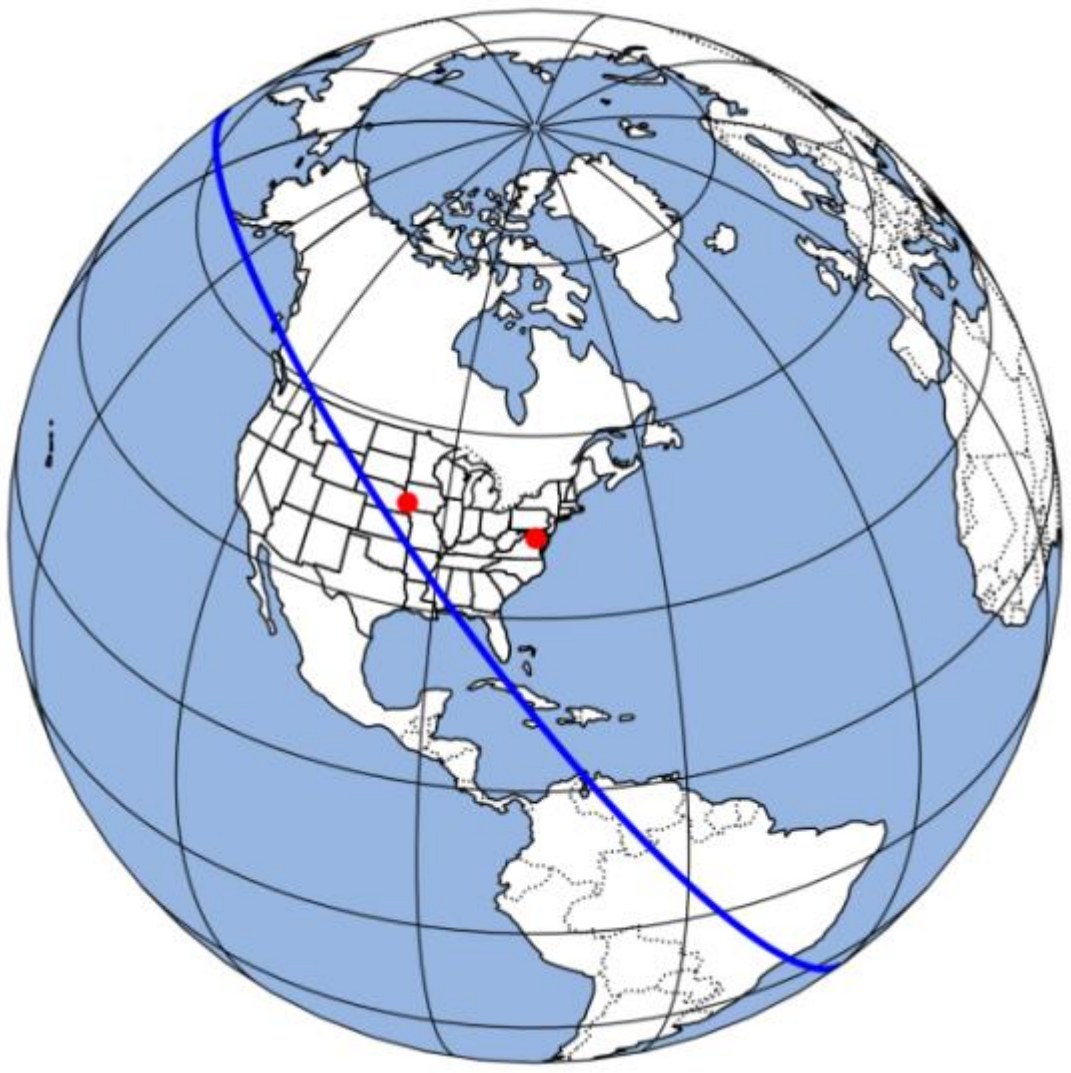

**Figure 13:** This figure shows the orbit of a launch similar to 2026-120A from Xichang, China, as the FOBS weapon passes over Omaha on its eighth orbit, 12.4 hours after launch.

Table 2 gives a summary of these attacks.

**Table 2:** A summary of the results of the calculations in the text for FOBS attacks launched by one and two SLVs. "Number of FOBS" is the number of FOBS weapons that could be carried to each target on the SLVs used in the launches.

| | | **Russian Attack Against:** | | **Chinese Attack Against:** | |
|---|---|---|---|---|---|
| | | **Washington** | **Omaha** | **Washington** | **Omaha** |
| **1 SLV** | Launch to first deorbit | 21 min | | 65 min | |
| | First deorbit to impact | 8.8 min | 8.5 min | 7.7 min | 7.3 min |
| | FOBS weapon mass | 370 kg | 1075 kg | 690 kg | 690 kg |
| | Number of FOBS | 9-10 | 3 | 8-9 | 2-3 |
| **2 SLV** | Launch to deorbit | 25 min | 9.6 hr | 4.6 hr | 12.4 hr |
| | Deorbit to impact | 2.7 min | 4.2 min | 3.7 min | 5.4 min |
| | FOBS weapon mass | 370 kg | 370 kg | 460 kg | 370-460 kg |
| | Number of FOBS | 18 | 6 | 10 | 6-8 |

## Testing

An important question is whether a country could use existing RVs for FOBS attacks, giving it a FOBS capability without conducting additional tests—either tests of existing RVs on a FOBS trajectory or development tests for a new RV. If testing is required for a country to have high confidence in a FOBS capability, the lack of such testing would give other countries some confidence that it does not have a reliable capability for such attacks. Similarly, because RV tests would likely be detected by other countries, tests that would provide validation of an RV for such attacks could provide evidence that the country is developing a FOBS capability.

To explore this question, we calculate the reentry heating of an RV on several trajectories and compare the conditions experienced on FOBS attacks with those on typical ballistic trajectories. As an RV is slowed by atmospheric drag during reentry its kinetic energy is converted to heating of the air around it, some fraction of which is transferred to the RV body. Because the initial energy of the RV is so large, delivering nuclear weapons by missiles required countries to develop ways to protect the warhead from that heat. Early RVs had blunt bodies with high drag coefficients so that much of the deceleration occurred at high altitudes where the thinner atmosphere transferred less of the heat to the RV body. But achieving highly accurate weapons required developing low-drag RVs that would pass through the atmosphere more quickly, which led to the development of ablative heatshields that use the phase change of the heatshield material to prevent significant heat from entering the RV body.

The two aspects of reentry heating we consider are the total heat the RV absorbs and the length of time during which significant heating occurs. Longer times result in more heat diffusing into the RV, raising the temperature in its interior, a process called "heat soak." Designing an RV to withstand greater heat absorption requires increasing the mass of ablative material in the heat shield. Dealing with longer times for heat soak requires increasing insulation in the heatshield to limit the diffusion of heat to the interior of the RV.

We calculate RV heating using empirical equations to estimate the heating on various trajectories and compare the heat absorbed and the heat soak time in each case (see Appendix).[44] Figure 14 shows the results of those calculations for a 10,000 km MET and two DTs launched on MMIII-like missiles, and four reentry trajectories from the FOBS analysis above. The DTs have apogees of 700 and 340 km compared to 1,475 km for the MET. The solid black curve assumes the FOBS RV reenters from 200 km altitude with no lateral displacement, so it impacts on the ground track of the FOBS, and with a deorbit booster providing $\Delta V$=1.5 km/s; this result is very similar to a reentry from 120 km. The black dashed and dotted curves are for the trajectories shown in Figures 8 and 10, which have significant lateral displacement and therefore long reentry paths within the atmosphere.

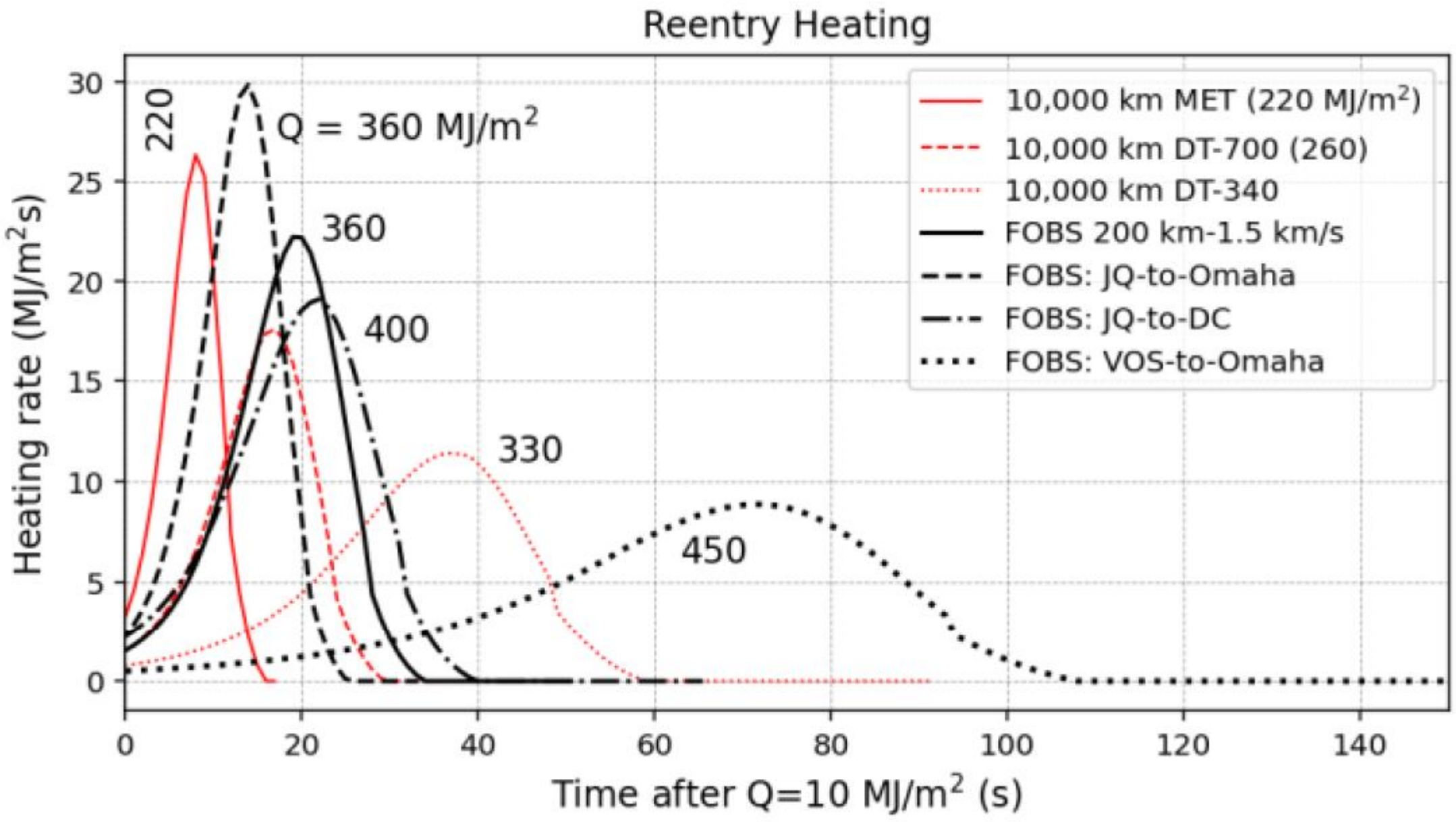

**Figure 14.** These curves show the reentry heating of RVs on different trajectories. The solid FOBS curve assumes deorbit from 200 km with $\Delta V$ = 1.5 km/s to a target on the ground track of the orbit. The other FOBS curves are for trajectories in Figures 8 and 10. Each curve is labeled by an estimate of the total heat absorbed per area (Q), which is the area under the curve. The zero of time for each curve is taken as the time when the RV has absorbed 10 MJ/m$^2$, and the curves stop at the time of impact. This allows comparison of the duration of heat soak in each case. The heating is estimated using empirical heating formulas assuming turbulent flow over the RV surface. The slope discontinuity in some of the curves is an artifact of the change in the empirical equations for high and low speeds.

The curves show the heat per area the RV absorbs over time, with t = 0 corresponding to the time at which each RV has absorbed 10 MJ/m$^2$, which we estimate will result in an RV surface temperature large enough to drive significant heat diffusion into the RV body.[45] The number beside each curve or in the legend gives the total heat absorbed per area at a point midway along the RV body from that time until impact. The heat soak time ($\tau$) is given by the time at which each curve ends, which corresponds to impact. These values are given in Table 3. The heat absorbed should be roughly proportional to the mass of heatshield material ablated during reentry, so the third column gives an estimate of the increase in heatshield mass that would be ablated relative to the MET.

Once the RV surface reaches the ablation temperature of the heatshield material (typically several thousand Kelvin), the heatshield will ablate but the surface temperature will remain at the ablation temperature. Analysis of the heat diffusion into the RV shows that the thickness of insulation $L$ needed to keep the inside surface of the insulation at the same temperature for different heat soak times $\tau$ varies as $L \sim \sqrt{\tau}$.[46] The fifth column in Table 3 gives the square root of the ratio of heat soak times for these trajectories to that of the MET, which is an estimate of the increased thickness of insulation required to keep the temperature at the inner surface of the insulation the same temperature in the two cases.

**Table 3.** A comparison of the heat absorbed and time of heat soak for the trajectories in Figure 14. The first FOBS row assumes deorbit from 200 km with $\Delta V$=1.5 km/s to a target on the ground track of the orbit. The other FOBS curves are for trajectories in Figures 8 and 10. The fifth column gives the square root of the ratio of heat soak times for each trajectory compared to the MET, which is proportional to the relative thickness of insulation required to keep the inner temperature the same as in the MET case.

| Trajectory | Heat absorbed Q ($MJ/m^2$) | $Q/Q_{MET}$ | Heat soak time ($\tau$) | $\sqrt{\tau/\tau_{MET}}$ |
|---|---|---|---|---|
| **MET 10k km Apogee = 1475 km** | 220 | 1.0 | 17 | 1.0 |
| **DT1 10k km Apogee = 700 km** | 260 | 1.2 | 42 | 1.6 |
| **DT2 10k km Apogee = 340 km** | 330 | 1.5 | 91 | 2.3 |
| **FOBS 200 km-1.5 km/s** | 360 | 1.6 | 51 | 1.7 |
| **FOBS JQ-to-Omaha** | 360 | 1.6 | 34 | 1.4 |
| **FOBS JQ-to-DC** | 400 | 1.8 | 71 | 2.0 |
| **FOBS VOS-to-Omaha** | 450 | 2.0 | 150 | 3.0 |

Based on publicly available information, neither China nor Russia are known to have conducted DT tests of long-range missiles. Russia conducted several reentry tests of FOBS weapons in the 1960s and may have useful information from those tests. The US conducted a DT test of the Trident II in March 2005 as part of the conventional Trident modification (CTM) program, which reportedly shortened the range from 7,400 km to 2,200 km with a flight time of 11 to 12 minutes.[47] Calculations show that this flight time would suggest a relatively high apogee—about 350 km. Because of the short range, which reduces the heating, this test would not be particularly useful for understanding the heating on a FOBS trajectory.

These numbers show that if the attacking country has not tested a long-range missile on a depressed trajectory, it may not have high confidence using the RV on FOBS trajectories, unless it believed its RVs were significantly overdesigned for the MET case, and had enough testing and high-fidelity computer modeling to extrapolate to these conditions. The heating of a FOBS weapon like the one deorbited from 200 km might be considered acceptable enough that a country would decide to use one in a time of crisis, if it believed it had few other options and was willing to accept the technical risk

The other potential testing issue could be to evaluate the accuracy of RVs on FOBS trajectories, which would travel a longer path within the atmosphere than a typical RV reentry and might use lift to improve the accuracy. This is likely only to be concern for FOBS trajectories that require significant lateral reach, and only against the hardest targets, like Raven Rock. This issue would also arise with RVs delivered on significantly depressed ballistic trajectories.

Negotiating a ban on testing needed to gain confidence in using FOBS could be a useful confidence-building measure, but may not prove decisive in a crisis.

**Conclusions**

FOBS are essentially depressed-trajectory missiles that can deliver warheads over long distances while remaining at much lower altitudes than a typical long-range ballistic missile. If successfully disguised as satellite launches they can also have very short warning times.

FOBS-like weapons do not represent a game-changing advance in technology or capability. They instead rely on well-developed technologies that have been used in space systems for decades, and that are currently being improved as countries develop space planes and hypersonic weapons. The conditions during reentry, however, may differ enough from those experienced by RVs tested on standard trajectories that a country may not have high confidence in using one as a FOBS without additional testing. The US would likely detect such testing and see it as evidence that a country was developing a FOBS capability.

An important cost of using FOBS compared to standard ballistic missile trajectories to deliver RVs is the significant reduction in payload mass. Putting a FOBS weapon into orbit requires the launch vehicle to accelerate its payload to higher velocity than is required to loft a long-range ballistic missile to 10,000 km range, reducing the missile's payload by more than 40 percent. In addition, part of the payload for a FOBS launch is taken up by the mass of the deorbit booster attached to the RV, which can be comparable to or significantly greater than the mass of the RV. Combining the payload reduction to allow a greater burnout speed with the fraction of payload consumed by a moderate-sized deorbit booster ($\Delta V = 1.5$ km/s), the effective RV payload for FOBS is less than one-third that of the same missile on a minimum-energy ballistic trajectory.

The most likely motivation for developing FOBS today is as a means of evading ballistic missile defenses. Doing so does not require flying a southern route to US territory as is often discussed. Flying FOBS weapons on orbits with altitudes of around 120 km would keep them at altitudes too low to be intercepted by US strategic missile defenses.[48] Whether it makes sense for a country to pursue FOBS for this purpose depends on the reliability of other means of penetrating missile defense. Countries that could build FOBS could also equip their missiles with decoys and other countermeasures to defeat missile defenses. The fact that launching a FOBS attack results in a significant payload reduction compared to launching a standard attack on a ballistic missile would make the option of using FOBS less attractive. Even if a country had confidence that it could use countermeasures to penetrate missile defenses, it might want to test an option that would make clear its ability to penetrate defenses clear to US officials as well.

A second potential use of FOBS is short-warning attacks. We considered three types of such attacks.

The first is an attack launched on ballistic missiles. As with ballistic missiles launched on normal ballistic trajectories, a FOBS attack using ballistic missiles would be identified almost immediately after launch. Because of the shorter flight path and higher speed, a long-range FOBS would have a flight time about 10 minutes shorter than a ballistic missile to the same range, and radar confirmation of the attack would come about 10 minutes later than for an MET. This reduction in warning time would stress decision-making timelines for the country under attack, but it is not short enough for an attacker to be confident that it could prevent a US retaliatory strike.

The other two types of attacks are delivered using either one or two satellite launch vehicles (SLVs). If an attacker was able to successfully disguise a FOBS attack as a normal satellite launch, warning times could be much shorter than those above. The number of warheads that could be launched in this way would be limited by the small number of launchers used in such an attack. As examples, we identified recent Russian and Chinese satellite launches that would allow delivery of at least two warheads against five key command-and-control targets (four in Washington and one in Omaha) using the same launch vehicle as the satellite launch and a similar orbit as the satellite it released.

In the case of an attack using a single SLV launch, there could be 8 to 9 minutes of warning from the start of the first deorbit burn to impact on the target, with impact 30 to 70 minutes after launch. An attack using two SLVs could reduce the time between deorbit burn and impact to 4 minutes, but impact would occur 10 to 12 hours after the first SLV launch. Attacks with 4 to 9 minutes of warning would likely prevent the relocation of the president and other key decision makers, and might prevent convening a decision conference or even notification that an attack is underway.

Disguising FOBS launches as satellite launches would be difficult. A NOTAM for a space launch is usually published two or three days in advance, making it difficult to execute a surprise, short-warning attack in a matter of hours without raising suspicions about the purpose of the launch. The US might announce that during a crisis it would shoot down satellites that were launched without a two-day notice. It is worth noting, however, that space launches are often postponed, leading to last-minute cancellations and short-notice activations. Rather than launch a disguised FOBS without notification, an attacker might issue a NOTAM early in a crisis and modify it repeatedly or on short notice, keeping the option open in case the crisis worsens. How this might play out in an actual crisis is unclear.

An attacker considering a decapitating first strike would want confidence in the capability and reliability of its FOBS weapons to successfully carry out such a strike. The reentry conditions for FOBS differ from those of RVs on standard trajectories, with some types of FOBS presenting more stressing conditions than those for which the RVs likely was designed. Although high-fidelity computer modeling might be sufficient to determine whether an RV could survive a FOBS reentry trajectory, a country might conclude that flight tests are required to have sufficient confidence. The performance of the FOBS deorbit booster could likely be tested in a way that looked like an orbit change of a satellite, but a test that included reentry of an RV would be difficult to disguise and would be evidence of a FOBS. Flight tests of such a weapon would be detected by the US.

A disguised FOBS decapitation attack would be extremely risky; an adversary would consider it only under the most dire circumstances. The uncertainties inherent in an attempted decapitation and the threat of US retaliation would be expected to dissuade an adversary from attempting such an attack in most cases. If, however, a crisis had escalated to the point that an adversary believed a US nuclear strike was likely, it might believe that a decapitating strike was the best option for reducing the damage it would sustain. Any retaliatory US nuclear strike must be authorized by the president or a successor.[49] An adversary might therefore reasonably believe that a decapitating attack that killed the president and most potential successors would prevent, substantially reduce, or significantly delay a retaliatory attack, even though much of the US arsenal would remain intact.

If a crisis had escalated to this point, however, an adversary could not be sure whether the US had relocated a presidential successor to a secure facility from which a strike could ordered (as is reportedly done for State of the Union addresses). The successor could be briefed on pre-planned response options in the event of a decapitation attack. To the extent the US sees the possibility of such a strike as a threat, it should consider what procedures could be put in place to make the prospect of a survivor with launch authority credible enough in a crisis to increase deterrence against such an attack.

These considerations illustrate the instabilities that could arise as a crisis escalates. They highlight the need for the countries to develop ways to prevent crises from escalating to this point and to be able to take steps to deescalate if they do. It could also be useful for the US, Russia, and China to agree to ban flight tests of RVs that would demonstrate a FOBS capability. These countries might also agree on strict prenotification of satellite launches during times of crisis. While such agreements would not preclude short-warning attacks, they could be useful confidence-building measures.

**Acknowledgements**

The authors thank Jonathan McDowell for providing feedback and orbital data, and for suggesting the attack option using two SLV launches.

## Appendix: The dynamics of deorbiting

We calculate the deorbit trajectory of a FOBS weapon by integrating the equations of motion, with the initial velocity in the horizontal direction at altitude $h$ and magnitude equal to the orbital $V_o$ velocity at $h$:

$$V_o = \sqrt{\frac{GM_e}{r_e + h}} \approx 7.91\left(1 - \frac{h}{2r_e}\right)\ km/s \qquad (A1)$$

where $GM_e = 3.987 \times 10^5$ km$^3$/s$^2$, $r_e = 6370$ km is the mean radius of the Earth, and $h$ is in km.

To deorbit, the FOBS weapon fires its booster, giving a thrust at an angle $\eta$ with respect to the direction of the velocity. The booster provides a velocity change $\Delta V$. We model the booster as having two identical stages, each providing $\Delta V/2$. Each stage has a propellant mass $M_p$ and a structural mass that is given as a fraction $s$ of the propellant mass, so that the total stage mass at the beginning of its burn is $M = (1+s)M_p$. Using two stages is more efficient and gives a smaller total booster mass for large values of $\Delta V$, and gives essentially the same result as a single stage with the same total mass at small values of $\Delta V$.

We assume the booster provides an average thrust $a$, so the total burntime of the booster is:

$$t_b = \frac{\Delta V}{a} \qquad (A2)$$

Each stage has an exhaust velocity $V_e$, given by $V_e = g_0 * I_{sp}$, where $I_{sp}$ is the specific impulse of the booster in units of s$^{-1}$ and $g_0$ is the gravitational acceleration at the surface of the Earth.

The mass $M_{FOBS}$ of the FOBS weapon for a given $\Delta V$ can be calculated using Equation 30 in this reference.[50] In this case it becomes:

$$M_{FOBS} = M_{RV}\,[F(\Delta V, V_e, s)\,]^2 \qquad (A3)$$

where $M_{RV}$ is the mass of the RV and:

$$F(\Delta V, V_e, s) = \left[(1+s)e^{-\Delta V/2V_e} - s\right]^{-1} \qquad (A4)$$

Using this notation, the masses of the two stages can be shown to be:

$$M_{stage1} = F(F-1)M_{RV} \qquad M_{stage2} = (F-1)M_{RV} \qquad (A5)$$

Specifying the values of $M_{RV}$, $\Delta V$, $V_e$, $s$, $a$, and the ballistic coefficient $\beta$ of the RV allows us to integrate the equations of motion (see below) to give the deorbit dynamics of the FOBS weapon. For this paper, we assume $s = 0.1$, $a = 8g$, $\beta = 10{,}000$ kg/m$^2$, and $V_e = 2.75$ km/s, which corresponds to a specific impulse of 281 s.

**Equations of motion**

We model missile flight over a spherical, non-rotating Earth using the coordinate system illustrated in Figure A1. The equations of motion below describe the magnitude of the velocity ($V$), its angle measured relative to the local horizontal ($\gamma$), the downrange angle over Earth ($\Psi$), the crossrange angle of the velocity ($\kappa$) relative to the initial downrange direction, and the crossrange angle over Earth ($\Omega$). The altitude is given by $h$, $\rho$ is the atmospheric density, $g$ is the gravitational acceleration at altitude $h$, $r_e$ is Earth's radius, $L/D$ is the vehicle's lift-to-drag ratio, and $\sigma$ is the angle of the lift force with respect to the vertical plane containing the velocity.[51] $C_dA$ is the product of the vehicle's drag coefficient and effective area appropriate to the body at different times in its trajectory. Atmospheric density is given by the 1976 US Standard Atmosphere.[52] Down-range and cross-range distances, measured over Earth's surface, are given by $r_e\Psi$ and $r_e\Omega$, respectively.

When the ballistic missile's boosters are burning during launch, the thrust $T$ will be nonzero, $\eta$ is the angle of the thrust with respect to the velocity in the vertical plane that contains $V$, $m$ is the mass of the missile boosters plus payload and decreases by the rate $dm/dt$ at which the propellant mass is burned, which given by Equation A12 where $I_{SP}$ is the specific impulse of the rocket motor and $g_0$ is the acceleration of gravity at Earth's surface. Those variables are similarly defined for the time during which the deorbit booster is burning. In this case, $\alpha$ determines the fraction of the deorbit booster thrust perpendicular to the vertical plane containing the velocity; $\alpha$ is assumed to be zero during launch of the missile.

Rotation of the Earth under the FOBS' orbit is calculated using the flight time of the historical satellite trajectories from launch to the location of the target, which gives the correct relationship of the orbit near the target at the time of impact.

$$\frac{dV}{dt} = \frac{T\cos\alpha\,\cos\eta}{m} - \frac{C_dA}{2m}\rho V^2 - g\sin\gamma \qquad (A6)$$

$$\frac{d\gamma}{dt} = \frac{V\cos\gamma}{r_e + h} + \frac{T\cos\alpha\,\sin\eta}{Vm} + \left({}^{L}/_{D}\right)\left(\frac{C_dA}{2m}\right)\rho V\cos\sigma - \frac{g}{V}\cos\gamma \qquad (A7)$$

$$\frac{d\kappa}{dt} = \frac{T\sin\alpha}{Vm} + \left({}^{L}/_{D}\right)\left(\frac{C_dA}{2m}\right)\frac{\rho V\sin\sigma}{\cos\gamma} \qquad (A8)$$

$$\frac{d\Psi}{dt} = \frac{V\cos\gamma\cos\kappa}{r_e + h} \qquad (A9)$$

$$\frac{d\Omega}{dt} = \frac{V\cos\gamma\sin\kappa}{r_e + h} \qquad (A10)$$

$$\frac{dh}{dt} = V\sin\gamma \qquad (A11)$$

$$\frac{dm}{dt} = \frac{T}{g_0 I_{SP}} \qquad (A12)$$

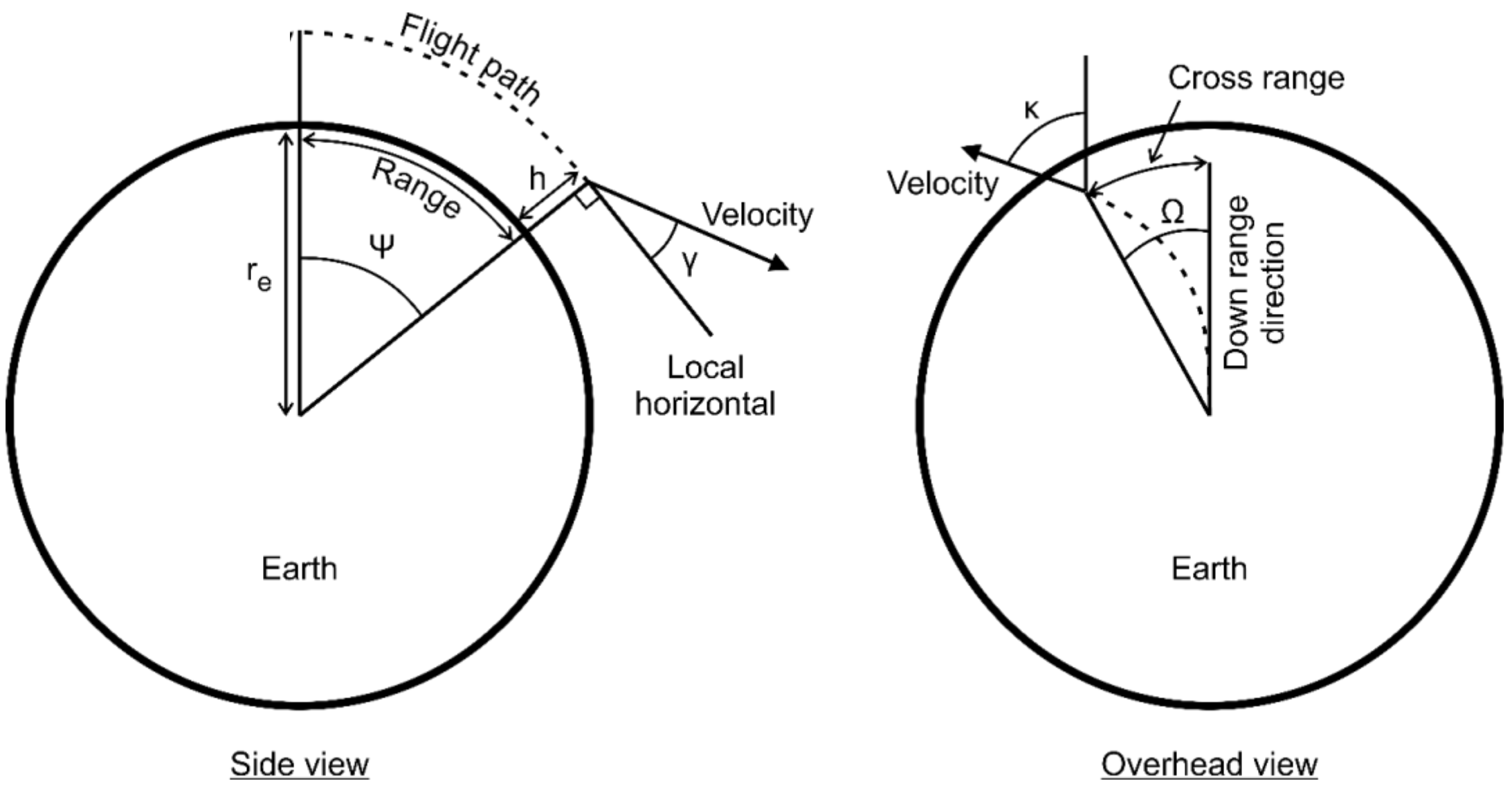

**Figure A1:** The coordinate system used for these calculations, shown from both side (left) and overhead (right) perspectives. Six trajectory variables are modelled: velocity (*v*), flight angles relative to the local horizontal (*γ*) and measured azimuthally from the down-range direction (*κ*), downrange angle over Earth (*Ψ*), crossrange angle over Earth (*Ω*), and altitude (*h*). Downrange and crossrange distances, measured over Earth's surface, are given by $r_e\Psi$ and $r_e\Omega$, respectively.

## Heating equations

We estimate heat transfer to the RV surface using standard phenomenological equations for heat transfer in turbulent hypersonic flow over a flat plate.[53] In the equations below, $dQ/dt$ is the heat flux per area in $J/m^2s$, $V$ is the velocity of the air relative to the vehicle in m/s, $h_w$ is the vehicle wall enthalpy per unit mass, $h_0$ is the stagnation enthalpy per unit mass, and $\rho$ is the atmospheric density in $kg/m^3$. The stagnation enthalpy is approximated as $h_0 = V^2/2 + 2.3\times10^5$ J/kg, and the wall enthalpy is approximated as $h_w = 1000T_w$ J/kg, where $T_w$ is the wall temperature in K.[54] We assume $T_w = 2{,}700$ K, which is the ablation temperature for refrasil phenolic resin used for heatshields.[55] $\theta$ is the angle between the vehicle surface and the freestream flow, and $x$ is the distance along the vehicle surface in meters from the leading edge; we assume $\theta = 8.5^\circ$ and $x = 0.75\ m$.

For $V > 4$ km/s:

$$\frac{dQ}{dt} = 2.2\times10^{-5}\left(\frac{(\cos\theta)^{2.08}(\sin\theta)^{1.6}}{x^{0.2}}\right)\left(1-\frac{1.11h_w}{h_0}\right)\rho^{0.8}V^{3.7} \qquad (A13)$$

and for $V \leq 4$ km/s:

$$\frac{dQ}{dt} = 3.89\times10^{-4}\left(\frac{(\cos\theta)^{1.78}(\sin\theta)^{1.6}}{x^{0.2}}\right)\left(1-\frac{1.11h_w}{h_0}\right)\left(\frac{556}{T_w}\right)^{0.25}\rho^{0.8}V^{3.37} \qquad (A14)$$

[1] Corresponding author: dcwright@mit.edu; 2 Windermere Lane, Arlington, MA 02476
[2] A history of the Soviet FOBS program is given in Asif A. Siddiqi, "The Soviet Fractional Orbiting Bombardment System (FOBS): A Short Technical History," *Quest,* 7:4, Spring 2000, p. 22-29, https://web.archive.org/web/20130401170310/https:/faculty.fordham.edu/siddiqi/writings/p15_siddiqi_quest_2000-04_fobs.pdf.
[3] United Nations Office for Disarmament Affairs, "Treaty on Principles Governing the Activities of States in the Exploration and Use of Outer Space, including the Moon and Other Celestial Bodies," https://treaties.unoda.org/t/outer_space.
[4] Siddiqi, "Soviet Fractional."
[5] The Soviet FOBS tests are listed in Mark Wade, "OGCh," *Astronautics,* http://www.astronautix.com/o/ogch.html and Jonathan McDowell, "5.6.9: FOBS," nd, https://www.planet4589.org/space/book/lv/engines/motorlist/fobs.html.
[6] Siddiqi, "Soviet Fractional."
[7] By the mid-1980s, the US also built two south-facing PAVE PAWS radars in Georgia and Texas, but they were deactivated in 1995.
[8] "Presidential Address to the Federal Assembly," 1 March 2018, http://en.kremlin.ru/events/president/news/56957. The video animation is available at https://www.c-span.org/program/international-telecasts/russian-president-vladimir-putin-state-of-the-nation-address/498311, at about 1:23:40.
[9] Demetri Sevastopulo and Kathrin Hille, "China tests new space capability with hypersonic missile," *Financial Times,* 16 October 2021, https://www.ft.com/content/ba0a3cde-719b-4040-93cb-a486e1f843fb; Jonathan McDowell, "Space Activities in 2021," 3 January 2022, Rev. 1.6, p. 51, https://www.planet4589.org/jcm/pubs/space/papers/annual/space21.pdf.
[10] Geoff Brumfiel, "New Chinese Space Plane Landed at Mysterious Air Base, Evidence Suggests," *All Things Considered,* 9 September 2020, https://www.npr.org/2020/09/09/911113352/new-chinese-space-plane-landed-at-mysterious-air-base-evidence-suggests.
[11] Siddiqi, "Soviet Fractional." Other advantages listed included unlimited flight range, short flight time, inability of the enemy to predict the FOBS target during its orbital flight, and high accuracy compared to ballistic missiles on a long trajectory.
[12] A.M. Sessler et al., *Countermeasures,* (Cambridge, MA: Union of Concerned Scientists and MIT Security Studies Program, April 2000), https://www.ucs.org/sites/default/files/2019-09/countermeasures.pdf.
[13] ODIN, "Poseidon Class (Kanyon Class) Russian Unmanned Underwater Vehicle," 12 March 2024, https://odin.t2com.army.mil/WEG/Asset/Poseidon_Class_(Kanyon_Class)_Russian_Unmanned_Underwater_Vehicle; Missile Defense Project, "Avangard," *Missile Threat*, Center for Strategic and International Studies, 23 April 2024, https://missilethreat.csis.org/missile/avangard/.
[14] Director Operational Test and Evaluation "Report in Support of National Missile Defense Deployment Readiness Review," 10 August 2000, p.14, http://www.bits.de/NRANEU/BMD/documents/coylereport.pdf; Sessler, *Countermeasures,* p. 28.
[15] The drag on an object moving with speed $V$ through atmosphere of density $\rho$ is proportional to $\rho V^2$ while the heating is proportional to $\rho V^3$. Soviet FOBS tests in the 1960s orbited at well under 200 km altitude; see Wade, "OGCh."
[16] The relative payload depends on the range to the target, which determines the speed of the ballistic missile; the altitude of the FOBS weapon, which determines its orbital speed; and the size of the deorbit booster. In this case, assuming the booster can deliver $\Delta V = 0.5$ or 1.5 km/s gives the relative number of RVs the missile could carry as 0.46 or 0.31, respectively.
[17] A hardened command post reportedly exists under the White House, but it is doubtful it would survive a direct strike by a high-yield nuclear warhead.
[18] Bruce Blair, "The U.S. Nuclear Launch and Decision Process," Global Zero, October 2019, https://www.globalzero.org/wp-content/uploads/2020/11/Full-LOWTimeline.pdf
[19] Similar concerns have been raised about the short flight time of missiles launched on depressed trajectories from submarines close to the US shores, but this would expose submarines to US antisubmarine warfare measures.
[20] Since the FOBS could have unlimited range, the relative payload penalty of the FOBS compared to an MET would be smaller for longer MET ranges.

[21] Areg Danagoulian, "Verification of the Outer Space Treaty with Cosmic Protons," 8 July 2026, https://arxiv.org/pdf/2512.20016.
[22] "Orbital launches by year and Country," *Space Stats*, 28 September 2026, https://spacestatsonline.com/launches/country/.
[23] Theresa Hitchens, "China's mysterious hypersonic test may take a page from DARPA's past," *Breaking Defense*, 24 November 2021, https://breakingdefense.com/2021/11/chinas-mysterious-hypersonic-test-may-take-a-page-from-darpas-past.
[24] James M. Acton, "Hypersonic Boost-Glide Weapons," *Science & Global Security* 23:3, 2015, https://scienceandglobalsecurity.org/archive/sgs23acton.pdf; David Wright and Cameron L. Tracy, "Hypersonic Weapons: Vulnerability to Missile Defenses and Comparison to MaRVs," *Science & Global Security*, vol. 31:3, 2023, https://scienceandglobalsecurity.org/archive/sgs31wright.pdf.
[25] Cameron L. Tracy and David Wright, "Modelling the Performance of Hypersonic Boost-Glide Missiles," *Science and Global Security*, vol. 28:3, 2021, http://scienceandglobalsecurity.org/archive/sgs28tracy.pdf.
[26] Tariq Malik, "Death of DARPA's superfast hypersonic glider explained," Space.com, 23 April 2012, https://www.space.com/15388-darpa-hypersonic-glider-demise-explained.html.
[27] Wright and Tracy, "Hypersonic Weapons: Vulnerability to Missile Defenses."
[28] Wright and Tracy, "Hypersonic Weapons: Vulnerability to Missile Defenses."
[29] McDowell, Jonathan C., 2026. General Catalog of Artificial Space Objects, Release 1.8.3, https://planet4589.org/space/gcat.
[30] To account for the rotation of the Earth under the orbit during the FOBS' flight, the launch is calculated as starting from a location west of the actual launch site by the distance the target will move east during that time, which is given by the flight time from launch to impact multiplied the rotational speed of the Earth at the latitude of the target. This gives the correct relationship of the orbit near the target at the time of impact.
[31] An object on this orbit would have a speed of 7.79 km/s and period of 88.3 minutes.
[32] Jonathan McDowell, personal communication; Anatoly Zak, "Soyuz launch Gonets trio, Skif-D, 24 April 2026, https://www.russianspaceweb.com/gonets-block18-skif.html.
[33] These times could be reduced by 30 to 60 seconds by using a deorbit booster for Omaha with $\Delta V$ = 4.5 km/s, which gives a FOBS weapon mass of 1350 kg. It would still be possible to still carry 2 of these and 8 FOBS weapons targeting DC on this SLV.
[34] EO-Atlas, "Roscosmos Soyuz 2," 16 June 2026, https://eo-atlas.org/launch_services/roscosmos-soyuz-2.
[35] Jonathan McDowell, personal communication.
[36] The State Council Information Service, People's Republic of China, "China launches Lijian-2 Y1 carrier rocket," *Xinhua*, 31 March 2026, http://english.scio.gov.cn/chinavoices/2026-03/31/content_118410990.html gives 12,000 kg for a 200-km orbit and 8,000 kg for a 500-kg SSO.
[37] Michael Listner, "FOBS, MOBS, and the reality of the Article IV nuclear weapons prohibition," *The Space Review,* 17 October 2022, https://www.thespacereview.com/article/4466/1.
[38] McDowell, General Catalog. The FOBS weapon would have a speed of 7.71 and period of 91.1 min
[39] Space Launch Now, "Angara-1.2," 9 July 2014, https://spacelaunchnow.app/vehicle/launch_vehicle/136/ gives 3,700 kg for "LEO" and 1,400 kg for "SSO."
[40] McDowell, General Catalog. The FOBS weapon would have a speed of 7.67 km/s and period of 92.4 min
[41] Space Launch Now, "Zhuque-2E Block 2," 14 May 2016, https://spacelaunchnow.app/vehicle/launch_vehicle/556/ gives 6,000 kg for "LEO" and 4,000 kg for "SSO."
[42] McDowell, General Catalog. The FOBS weapon would have a speed of 7.65 km/s and period of 93.4 min
[43] Federal Aviation Administration, "The Annual Compendium of Commercial Space Transportation: 2018," January 2018, https://www.faa.gov/about/office_org/headquarters_offices/ast/media/2018_AST_Compendium.pdf; China Great Wall Industry Corporation, "LM-2D," 1 April 2010, https://web.archive.org/web/20170721043418/http:/www.cgwic.com/LaunchServices/LaunchVehicle/LM2D.html. These sources give 1,300 kg into a 200-km orbit, and 1,300 into a 645-km SSO.
[44] Our calculations assume turbulent air flow over the RV, which should be true for the times during which most of the heat is absorbed.
[45] This heat absorption will increase the RV surface temperature by of many hundred Kelvin. The penetration depth for the heatshield a time $t$ after a temperature change $\Delta T$ is applied to the surface is $\delta \sim \sqrt{\kappa t/\rho C_p}$. If heat per area $Q$ is absorbed in a region of thickness $\delta$, the temperature rise (in K) of the volume will be $\Delta T \sim Q/\delta\rho C_P$ $\sim 1x10^{-3} Q/\sqrt{t}$, where the last equality is for refrasil phenolic, with $\kappa$ = 0.5 W/mK, $\rho$ = 1,632 kg/m$^3$, $C_P$ = 1,174 J/kgK. For $Q$ = 1 MJ much of this heat is absorbed in 20-60 s for these trajectories, giving $\delta \sim 0.2 - 0.4\ cm,$

$\Delta T \sim 1{,}300 - 2{,}240\ K$. (See C.J. Katsikas, G.K. Castle, and J.S. Higgins, *Ablation Handbook,* AVCO Corporation Technical Report AFML-TR-66-262, September 1966, p. 251, and John Lienhard V and John Lienhard IV, *A Heat Transfer Textbook (6th edition),* Phlogiston Press, Cambridge, MA, 2024, p. 224, https://bpb-us-e1.wpmucdn.com/sites.mit.edu/dist/6/1810/files/2024/04/AHTTv600.pdf.)

[46] See, for example, Daniel W. Mackowski, "Conduction Heat Transfer: Notes for MECH 7210, " Mechanical Engineering Department, Auburn University, https://www.professores.uff.br/diomarcesarlobao/wp-content/uploads/sites/85/2017/09/condbook.pdf, starting on page 48.

[47] Norman Polmar, "U.S. Navy: Conventional Trident on Hold," *US Naval Institute Proceedings*, November 2006, https://www.usni.org/magazines/proceedings/2006/november/us-navy-conventional-trident-hold.
Hans Kristensen, "Global Strike: A Chronology of the Pentagon's New Offensive Strike Plan," Federation of American Scientists, 15 March 2006, https://web.archive.org/web/20080306064659/http://www.fas.org/ssp/docs/GlobalStrikeReport.pdf.

[48] The US THAAD missile defense system is designed to intercept shorter range missiles at this altitude, but the speed of FOBS weapons would be much higher than it is designed to engage. Moreover, proposed space-based boost-phase interceptors would not be able to reach current-generation missiles, which have burn times of about three minutes, operating at such low altitudes, even assuming the kill vehicle's sensors were not blinded by heating.

[49] In the 1990s, the United States publicly acknowledged that "coded control devices" had been added to ballistic missile submarines, so that, as with nuclear weapons on land-based missiles and bombers, they require an externally transmitted authorization code to launch; William Perry, *Annual Report to the President and the Congress*, Department of Defense, February 1995, p. 91, https://www.dmi-ida.org/knowledge-base-detail/perry-95-annual-report-to-the-president-and-the-congress; U.S. Arms Control and Disarmament Agency (ACDA), U.S. Commitment to the Treaty on the Non-Proliferation of Nuclear Weapons, ACDA factsheet, 15 October 1997, https://nuke.fas.org/control/npt/docs/commnpt.htm; Office of the Deputy Assistant Secretary of Defense for Nuclear Matters, "Chapter 8: Nuclear Surety," *The Nuclear Matters Handbook 2020 (revised),* nd, https://www.acq.osd.mil/ncbdp/nm/NMHB2020rev/chapters/chapter8.html.

[50] David Wright, "Basic Equations of Space-Based Missile Defense," Laboratory for Nuclear Security and Policy, 3 June 2026, https://lnsp.mit.edu/s/Wright-Basic-eqns-of-space-based-MD.pdf.

[51] Ping Lu, Stephen Forbes, and Morgan Baldwin, "Gliding Guidance of High L/D Hypersonic Vehicles," AIAA Guidance, Navigation, and Control (GNC) Conference (2013); Luhua Liu, Jianwen Zhu, Guojian Tang, and Weimin Bao, "Diving Guidance via Feedback Linearization and Sliding Mode Control," *Aerospace Science and Technology* 41 (2015): 16-23; JianHua Wang, LuHua Liu, and GuoJian Tang, "Guidance and Control System Design for Hypersonic Vehicles in Dive Phase," *Aerospace Science and Technology* 53 (2016): 47-60; Shili Tan, Humin Lei, and Tao Liu, "Optimal Maneuver Trajectory for Hypersonic Missiles in Dive Phase Using Inverted Flight," *IEEE Access* 7 (2019): 63493-63503.

[52] National Oceanic and Atmospheric Administration, National Aeronautics and Space Administration, and United States Air Force, *US Standard Atmosphere, 1976* (Washington, DC: US Government Printing Office, 1976).

[53] Michael E. Tauber, Gene P. Menees, and Henry G. Adelman "Aerothermodynamics of Transatmospheric Vehicles," *Journal of Aircraft* 24 (1987): 594-602; John D. Anderson, *Hypersonic and High-Temperature Gas Dynamics*, 2nd ed. (Reston, VA: American Institute of Aeronautics and Astronautics, 2006), 349-350.

[54] John J. Martin, *Atmospheric Reentry* (Englewood Cliffs, NJ: Prentice-Hall, 1966), 16, 114.

[55] Katsikas, *Ablation Handbook.*